# Microscale optoelectronic synapses with switchable photocurrent from halide perovskite

Jeroen J. de Boer[1], Agustin O. Alvarez[1], Moritz C. Schmidt[1], Dimitrios Sitaridis [1], Bruno Ehrler[1,*]

[1]LMPV-Sustainable Energy Department, AMOLF, 1098 XG, Amsterdam, the Netherlands
*Correspondence: b.ehrler@amolf.nl

## Abstract

Efficient visual data processing by neuromorphic networks requires volatile artificial synapses that detect and process light inputs, ideally in the same device. Here, we demonstrate microscale back-contacted optoelectronic halide perovskite artificial synapses that leverage ion migration induced by a bias voltage to modulate their photocurrent. The photocurrent changes are due to the accumulation of mobile ions, which induces a transient electric field in the perovskite. The photocurrent changes are volatile, decaying on the order of seconds. The photocurrent changes can be controlled by both the applied voltage and illumination. The symmetric device supports changing of the photocurrent polarity, switching between inhibitory and exhibitory functioning. The photocurrent can be updated by spike-timing-dependent plasticity (STDP)-learning rules inspired by biology. We show with simulations how this could be exploited as an attention mechanism in a neuromorphic detector. Our fabrication procedure is compatible with high-density integration with CMOS and memristive neuromorphic networks for energy-efficient visual data processing inspired by the brain.

## Introduction

Rapid developments in the field of artificial intelligence (AI) have led to impressive performance of neural networks over a broad range of tasks, such as natural language processing,[1,2] image recognition,[3,4] and protein folding prediction.[5] However, the increase in the capabilities of neural networks has come at the price of exponentially increasing energy consumption.[6] Neuromorphic computing offers a more energy-efficient alternative to neural networks run on classical computers.[7] In neuromorphic computing, electronic analogs to biological neurons and synapses mimic highly energy-efficient biological neural networks. Similar to their biological counterpart, neuromorphic artificial synapses process and store information within the network by changing the synaptic connection strength between neurons, typically through a variable resistance. This can be implemented with memristive devices, which have a resistance that can be varied by applying a bias voltage.[8] Volatile memristive devices, of which the resistance change decays to a steady-state high-resistive state over time, are particularly well-suited to mimic brain-like filtering and processing of sensory information.[9,10] For this application, the volatility of the devices ensures that signals that occur at different points in time can be distinguished. When applied as a filter, the output intensity of volatile devices changes depending on recent input, for example, allowing them to function as bandpass filters.[9] Volatile devices can also be implemented for short-term working memory. Information relevant to a task, such as speech recognition or recalling a recently detected object, is stored for a short time and automatically forgotten after the task is completed.[11,12] Here, the volatility of the memristive devices prevents the storage of information that is no longer relevant to the network.

Halide perovskites are an emerging class of semiconducting materials for neuromorphic devices. Highly mobile ionic defects in these materials readily cause hysteresis, which has been leveraged to fabricate energy-efficient artificial synapses,[13–15] and, more recently, neurons.[16] Hysteresis typically occurs on the hundreds of milliseconds to seconds time-scales,[17,18] ideal for volatile synapses. Moreover, their easy solution-processability allows facile deposition, even on flexible substrates.[19,20] Halide perovskites are also excellent light absorbers, with a band gap that can be tuned by incorporating different halides.[21] A notable property of halide perovskites is that the ionic mobility is coupled to light absorption, with higher ionic mobilities under more intense illumination conditions.[22] This interplay of ionic mobility and the photogenerated charge carriers might therefore enable the use of halide perovskites for simultaneous detection and processing of visual input. This application would further enhance the energy efficiency and scalability of neuromorphic networks.[23]

Although first macroscale implementations of optoelectronic halide perovskite synapses that can process light pulses have shown impressive energy[14,24] and light-detection efficiencies,[25] their scalability remains challenging. This is due to complex device architectures requiring multiple active layers,[26,27] sometimes with a third gate electrode,[25,28,29] and the high solubility of halide perovskites in polar solvents,[30] which makes microfabrication with lithography difficult.

Here, we adapt the lithographic procedure we recently developed for all-electronic halide perovskite synapses and neurons[13,16] to fabricate volatile halide perovskite optoelectronic synapses on the microscale. Applying a voltage pulse to the synapse causes subsequent volatile photocurrent enhancement that decays over approximately 5 seconds. From transient photocurrent measurements and drift-diffusion simulations, we deduce that the mechanism is the accumulation of mobile ions by the bias voltage, which induces an electric field in the perovskite. We postulate that this improves the extraction of photogenerated charge carriers under illumination, resulting in a volatile photocurrent enhancement. We demonstrate that the photocurrent enhancement is more significant for higher applied bias voltages and light intensities during the voltage. We attribute this dependence to the higher ionic mobility under illumination and therefore more significant ionic accumulation for higher applied voltages. We show that the photocurrent polarity can be both positive and negative depending on the applied voltage, a unique feature of our device that enables facile switching from excitatory to inhibitory functioning. We fabricate devices that are sensitive to different parts of the visible spectrum by incorporating $MAPbI_3$, $FAPbBr_3$, and $FAPb(I_{0.5}Br_{0.5})_3$ perovskites, and show that the photocurrent changes generalize over all of these perovskite layers. Finally, we show STDP learning by the synapse and simulate how this could be used for a neuromorphic detector employing an attention mechanism. The easily scalable device structure and mild conditions during the fabrication process allow the implementation of this detector with existing neuromorphic chips.

# Results and Discussion

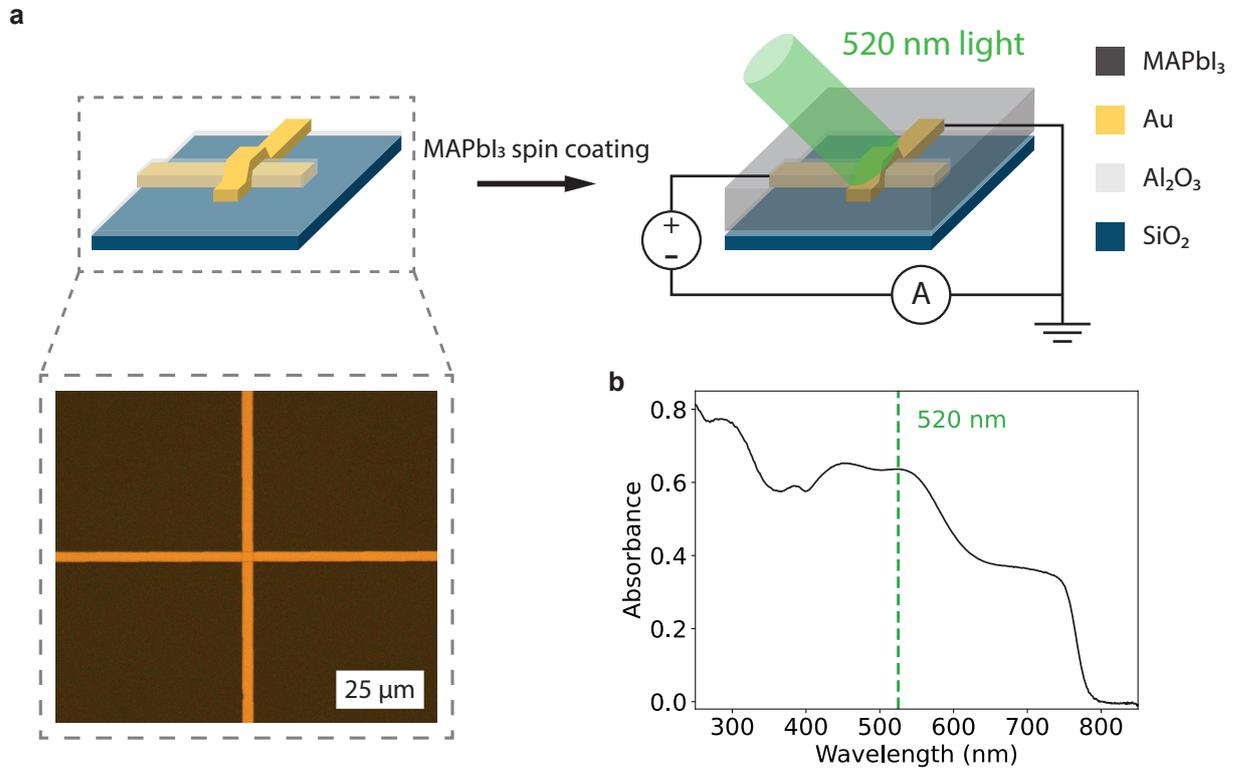

*Figure 1. The volatile halide perovskite optoelectronic synapse.* **(a)** *Schematic drawing of the halide perovskite optoelectronic synapse. After microfabrication of the 2.5 μm wide gold electrodes, the MAPbI$_3$ layer is spin-coated on the substrate. The device is connected to a source-measure unit (SMU) to apply voltages and measure current, and light is supplied as an additional input.* **(b)** *UV-Vis absorption spectrum of the MAPbI$_3$ film on a quartz substrate. The halide perovskite layer absorbs light over a broad range of wavelengths.*

Figure 1a shows a schematic and optical microscopy image of our microscale halide perovskite volatile optoelectronic synapse (MPOS). Two 2.5 μm wide gold electrodes form a cross-point device that sandwiches an approximately 15 nm ALD-deposited Al$_2$O$_3$ layer. We spin-coat the MAPbI$_3$ active layer over the electrodes in the final step to prevent degradation during lithography. Voltages and currents can be applied and measured between the top and bottom electrodes of the device. Importantly, the back-contact architecture also allows for the illumination of the MAPbI$_3$ layer without unwanted reflections off the gold electrodes.

Figure 1b shows that the MAPbI$_3$ layer we employ efficiently absorbs light with wavelengths below 780 nm. The absorption for 520 nm light, used in later measurements, is highlighted with a green dotted line. The broad absorption range enables the MPOS to process input light signals over the whole visible range and into the UV. This makes it more versatile than optoelectronic synapses based on materials that can only process UV-light inputs due to their limited absorption in the visible spectrum.[31,32]

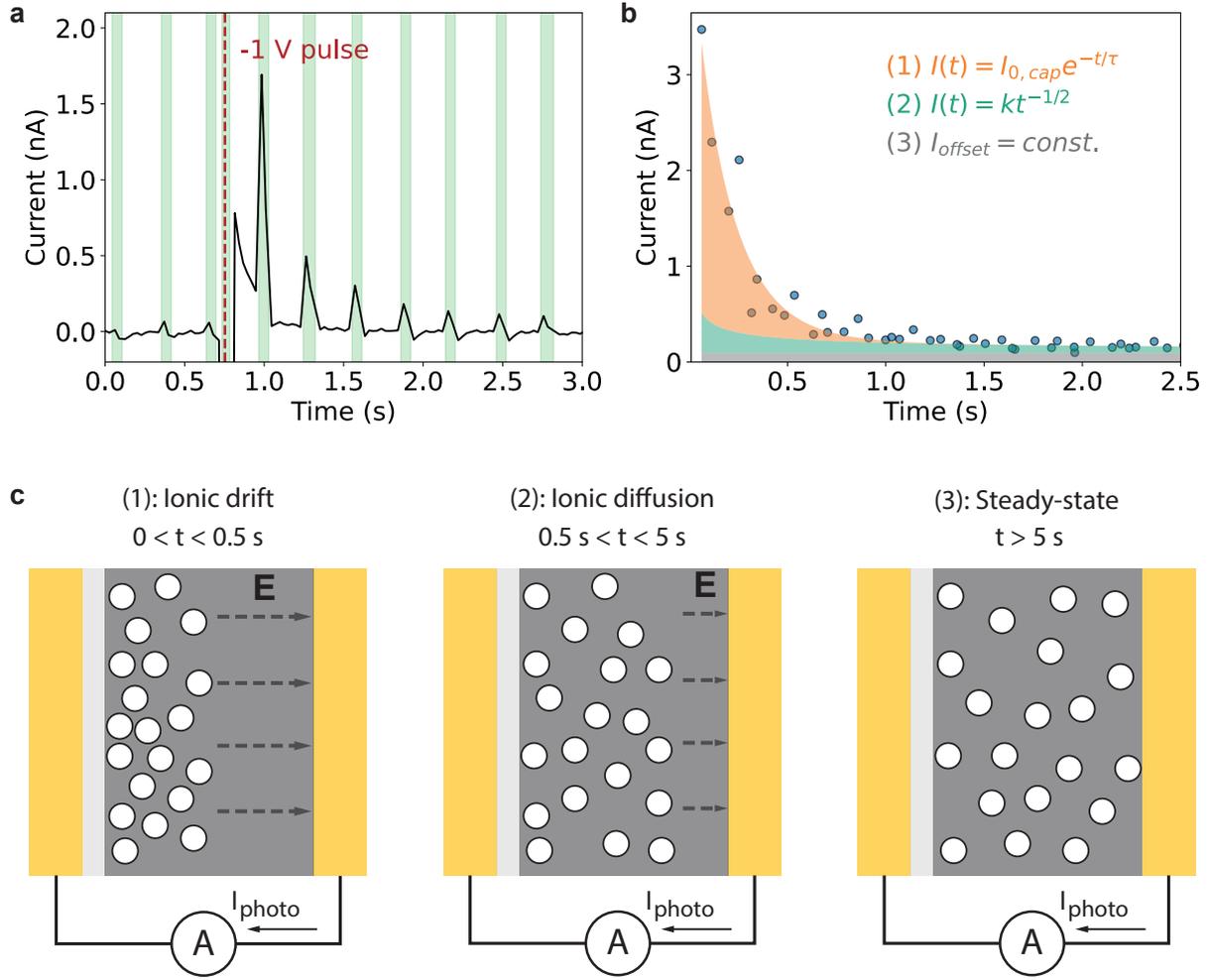

*Figure 2. Setting and reading out of the state of the optoelectronic synapse. A small initial photocurrent is read out when the device is illuminated with a green LED, indicated by the green regions in **(a)**. The photocurrent is increased after applying a -1 V pulse, indicated with the red dotted line, and then decays over several seconds. **(b)** Photocurrent after applying the -1 V pulse, data from five measurements. Data fitted with drift, diffusion, and a constant offset current: $I(t) = I_{0,cap}e^{-t/\tau} + kt^{-1/2} + I_{offset}$. **(c)** Schematic of the proposed mechanism of the transient photocurrent response. The applied voltage pulse accumulates halide vacancies (white circles) at the cathode, resulting in an electric field inside the perovskite. At 0 < t < 0.5 seconds, the vacancies redistribute by a drift process, resulting in an exponential decrease of the induced electric field and hence the measured photocurrent, as captured by equation (1) in **(b)**. For later times, 0.5 < t < 5 seconds, the vacancies redistribute predominantly due to diffusion, resulting in a transient current response according to (2). After approximately 5 seconds, further decay becomes negligible and only the constant offset current (3) is measured.*

Figure 2a shows an example measurement of the synapse. Initially, pulsing a 520 nm LED gives a small photocurrent of tens of picoamps. After applying a -1 V pulse to the device under illumination, an initial increase of the current is measured in the dark. This current decreases exponentially. When a light pulse is applied during this decay process, a strongly increased photocurrent of 1.7 nA is measured. The enhanced photocurrent then decays over time as measured with each successive light pulse.

We analyze the photocurrent by repeating the measurement in Figure 2a five times. The measured photocurrents are plotted over time in Figure 2b, where t=0 indicates the time where the voltage pulse is removed. To fit the photocurrent decay over time, we assume a decay due to a combined drift and diffusion process, with a constant offset current: $I(t) = I_{0,cap}e^{-t/\tau} + kt^{-1/2} + I_{offset}$. This time-dependence of the current follows from the linear regions in plots of the transient photocurrent on the $t^{-1/2}$ and semi-log scale in Figure S1a and b, respectively. The obtained fitting parameters are given in Table 1. The stacked plot in Figure 2b of both contributions demonstrates an initial decay predominantly due to the exponential $I_{0,cap}e^{-t/\tau}$ term (drift). After approximately 0.5 seconds, the photocurrent decays predominantly according to the $kt^{-1/2}$ contribution (diffusion).

I-V sweeps of the device in the dark and under constant illumination are given in Figure S2. Both sweeps show a combined capacitive and resistive response without obvious signs of resistance changes within each scan, as can be seen in perovskite memristive devices.[14] We, therefore, exclude resistance changes as the origin of the photocurrent enhancement. Instead, we propose the model presented schematically in Figure 2c. During the applied voltage pulse, the mobile iodide vacancies in the perovskite layer accumulate at the interface between the perovskite and the cathode.[33] This causes a large potential drop close to the perovskite-cathode interface, resulting in a screening of the electric field in the perovskite bulk. After the voltage is removed, accumulated iodide vacancies induce an electric field inside the perovskite, as indicated in the left panel of Figure 2c. This field causes extraction of photogenerated charge carriers.[34] On a timescale of hundreds of milliseconds, the electric field induces the drift of the positively charged halide vacancies away from the cathode. The resulting exponential decrease in the magnitude of the electric field then results in a proportional decay of the photocurrent. After approximately 0.5 seconds, the halide vacancies have partially redistributed in the perovskite layer, as indicated by the middle panel in Figure 2c. Now the electric field inside the perovskite is smaller and the ions further re-distribute by a diffusion-limited process, which follows a $t^{-1/2}$ time dependence.[35] This competition between ionic drift and diffusion currents in memristive devices is well-known for other materials.[36] After the redistribution of the halide vacancies, the device reaches a steady state where only the original small offset current is measured, as shown in the right panel in Figure 2c. The origin of this offset current is likely a small electronic asymmetry in the device, e.g. because one electrode is covered with the $Al_2O_3$ layer, or it could be due to a defective perovskite-electrode interface. The decay over hundreds of milliseconds to seconds we measure here is in agreement with previous work on ion migration in halide perovskites.[17,18] Our proposed mechanism is supported by drift-diffusion simulations, given in Figure S3, which show the accumulation of halide vacancies and resulting build-up of an electric field in the device by a -1 V pulse. The simulations show a redistribution of the vacancies and consequential decay of the electric field over similar timescales as in Figure 2b. Figure S4 demonstrates that there is also an ionic current contribution to the total current measured when a light pulse is applied. However, this contribution is only minor, in line with previous work on similar halide perovskite devices.[17]

Table 1. Fitting parameters obtained for the data in Figure 2b. The offset current $I_{offset}$ was determined by taking the mean of the photocurrents measured before application of the -1 V pulse. Errors indicate one standard deviation.

| Fitting parameter | Value |
| --- | --- |
| $I_{0,cap}$ | $3.9 \pm 0.2\ nA$ |
| $\tau$ | $0.19 \pm 0.01\ s$ |
| $k$ | $0.11 \pm 0.02\ nA\sqrt{s}$ |
| $I_{offset}$ | $0.095 \pm 0.014\ nA$ |

Figure S5 demonstrates that applying a voltage pulse of opposite polarity results in a photocurrent enhancement with a similar magnitude but opposite polarity, following the same combined exponential and power-law decay. This further supports our proposed model, which predicts that the iodide vacancies accumulate at opposite electrodes for positive and negative voltages. The resulting electric fields and photocurrents are, therefore, also of opposite polarity. The ability to tune the polarity of the photocurrent, i.e. the synaptic weight, is a notable feature of the MPOS. Typically, the output current polarity of an artificial synapse is bound to that of a read voltage.[37] Excitatory and inhibitory functioning of a synaptic device in a neuromorphic network are therefore set by the circuitry and applied read voltage. By contrast, the weights of the MPOS can be switched to either positive or negative values for the same input optical pulse. This allows facile switching between inhibitory and excitatory functioning by the same synapse.

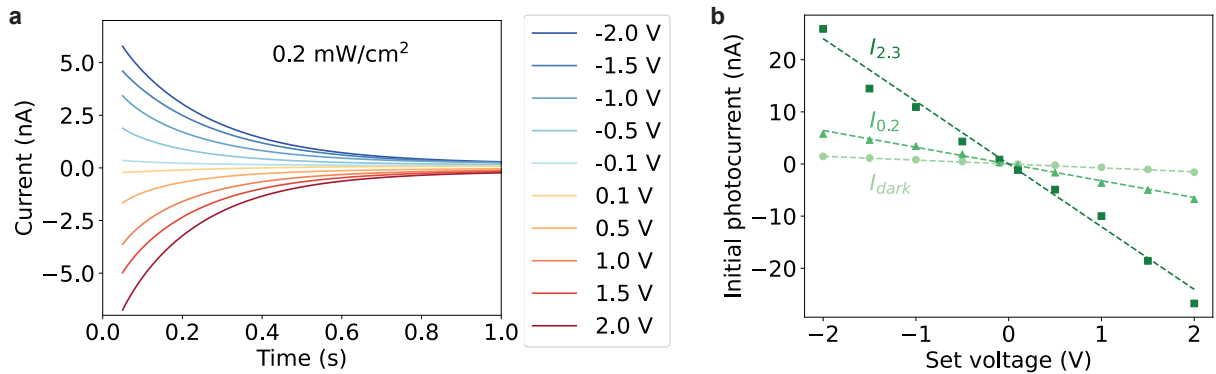

*Figure 3. Modulation of the photocurrent change with different voltages and light intensities.* ***(a)*** *Fits to transient photocurrent measurements with applied voltage pulses ranging from -2.0 to +2.0 V. All with the same light intensity of 0.2 mW/cm² during the voltage pulse. Larger photocurrent changes are measured for larger voltage amplitudes.* ***(b)*** *Comparison of the initial photocurrents, i.e. the photocurrents at 0.05 seconds from the transient current fits, for different applied voltages and light intensities of 2.3 mW/cm² ($I_{2.3}$, squares), 0.2 mW/cm² ($I_{0.2}$, triangles), and no light ($I_{dark}$, circles) during the voltage pulse. Linear fits to the initial photocurrents yield steeper slopes of -12.0 and -3.2 nA/V for $I_{2.3}$ and $I_{0.2}$, respectively, compared to the slope of -0.7 nA/V for $I_{dark}$.*

Figure 3a demonstrates that the magnitude of the synaptic weight can be tuned by varying the magnitude of the applied voltage. Larger voltage amplitudes, which are expected to cause the accumulation of a larger number of iodide vacancies, cause a larger photocurrent enhancement. In all cases, the photocurrent decays according to the combined drift and diffusion process.

The initial photocurrents at 0.05 seconds in Figure 3a are plotted in Figure 3b. From this figure, it follows that the magnitude of the photocurrent depends linearly on the applied voltage. The linear dependence can be explained by ion drift to the electrodes when applying the bias voltage. When the voltage is applied, an electric field is built up in the device according to $V(t) = V_{\sup}(1 - e^{-t/\tau})$.[38,39] Assuming the same characteristic time $\tau$ and using the same duration for each applied voltage pulse t, the voltage in the device increases linearly with the input voltage $V_{\sup}$. The linearity of the photocurrent change with respect to the input voltage makes weight changes of the MPOS with different voltages easily predictable. This predictability is important for the reliable training of neuromorphic networks.[40]

Figure 3b shows that the magnitude of the photocurrent is also altered by the illumination intensity during the voltage pulse. A higher irradiance of 2.3 mW/cm² ($I_{2.3}$) causes a larger photocurrent change compared to a lower irradiance of 0.2 mW/cm² ($I_{0.2}$). The photocurrent change is the smallest when the device is not illuminated during the voltage pulse ($I_{dark}$). The transient photocurrents over the whole voltage range for $I_{2.3}$ and $I_{dark}$ are given in Figure S6. These measurements show a similar photocurrent decay initially dominated by a drift process, which transitions to a diffusion-limited current decay at later times. The greater magnitude of the photocurrent changes at higher illumination intensities can be explained by the higher ionic mobility in halide perovskites under illumination.[22] A higher ionic conductivity during the applied voltage pulse causes a more rapid build-up of the electric field, i.e. accumulation of halide vacancies. The larger electric field in the device leads to a larger photocurrent when light pulses are applied.

The trends we observe for the photocurrent enhancement are not limited to 520 nm light excitation. Figure S7 shows the same measurement as in Figure 3, repeated with 450 and 620 nm light. These light sources, matched by photon flux to the 520 nm source, give similar photocurrent changes that are linear with the input voltage, demonstrating that the synapse can be operated with wavelengths ranging over the visible spectrum.

The photocurrent enhancement is also generalizable over different perovskites. As an example, we fabricated the same optoelectronic synapses with $FAPbBr_3$ and $FAPb(I_{0.5}Br_{0.5})_3$ active layers. Figure S8a and b show the absorption spectra of $FAPbBr_3$ and $FAPb(I_{0.5}Br_{0.5})_3$, respectively. The absorption onset of the perovskite layer shifts to shorter wavelengths for higher bromide contents. The -1 V measurements in Figure S8c and d (520 nm illumination) and Figure S9a and b (450 nm illumination), as well as the corresponding input voltage and illumination intensity sweeps in Figure S8e and f and Figure S9c and d show the same trends for these perovskite compositions as for the $MAPbI_3$ films. These results show that we can exploit the easily tunable band gap of halide perovskites to fabricate optoelectronic synapses that are only sensitive to specific wavelengths of light.

An important feature of the MPOS is that the synaptic weight of the device is changed with electronic pulses while it is read out with light pulses. Typically, both occur with the same type of input, i.e., all electronically or all optically, which can result in accidental weight changes during read-out.[12,41] Three-terminal synaptic transistors prevent this issue by using two lateral electrodes for read-out and a third gate electrode for weight updates.[42,43] However, this comes at the cost of higher device complexity and size and, therefore, scalability. Using optical signals for read-out and electronic signals for weight updates prevents accidental weight changes while maintaining the scalability of two-terminal devices.

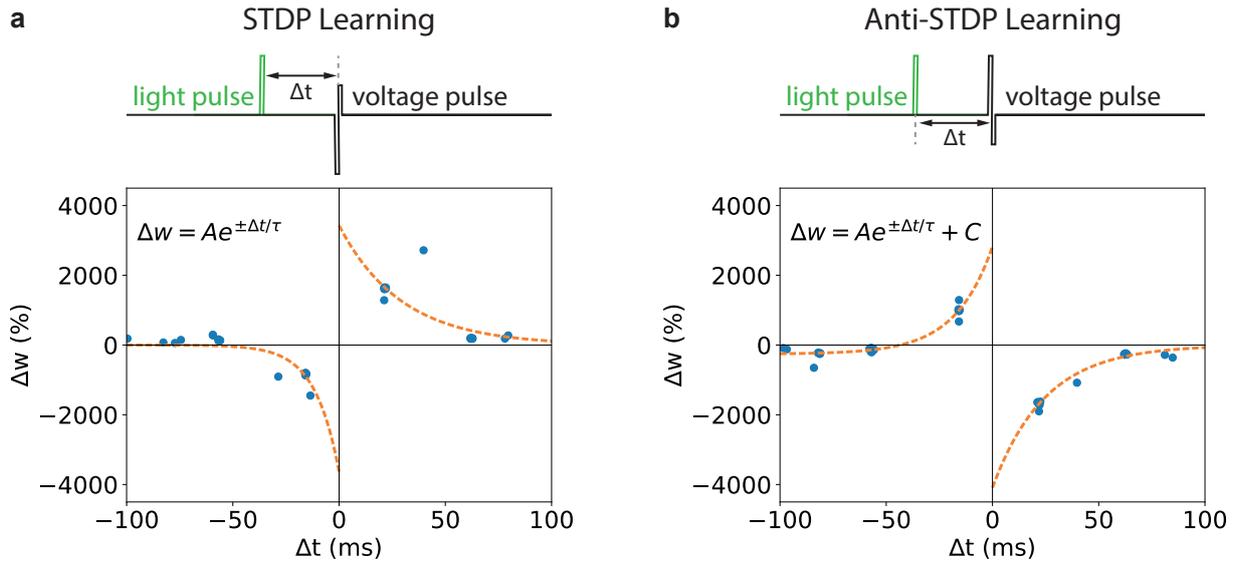

*Figure 4. STDP learning rules applied to the synapse. Applying a -1 to +0.5 V pulse results in STDP learning in **(a)**, while a +1 V to -0.5 V pulse leads to anti-STDP learning in **(b)**. The weight updates were fit to exponential decay or growth, with the addition of an offset in **(b)** to prevent overestimation of the exponential growth for -Δt due to the negative data points at Δt < -50 ms. Highly symmetric weight changes are obtained for STDP and anti-STDP learning, as well as for +Δt and -Δt within both learning rules.*

To demonstrate learning by the MPOS, we perform STDP measurements. A classic example of STDP-like learning in neuromorphic networks is inspired by Pavlovian conditioning, typically demonstrated in simple two-input, one-output networks. The first input ("sight of food") and the output neuron ("salivation") are initially correlated, meaning they are connected by a synapse with a high weight. In all-electronic neuromorphic networks, this is implemented as a high conductance of the synapse. The second input ("ringing of a bell"), on the other hand, is initially not correlated to the same output, which is implemented as a low synaptic weight, or low conductance of the second synapse. Due to the difference in the synaptic weights, initially, inputs through the high-conductance synapse cause spiking by the output neuron, while inputs through the low-conductance synapse do not. However, by presenting both inputs simultaneously, this simple network can learn to associate the two inputs through STDP. In STDP, back-propagating pulses are generated by the neuron as it spikes due to inputs through the high-conductance synapse. Overlap with simultaneous inputs through the low-conductance synapse results in a voltage drop that is large enough to increase its weight. After learning, presenting only the second input is enough to cause spiking of the output neuron, i.e. ringing the bell causes salivation by the dog.[44–46] STDP learning is commonly employed in large neuromorphic networks,[47] for example, for the recognition of complex patterns, such as handwritten digits.[48]

Figure 4 shows optoelectronic STDP measurements of the MPOS. Contrary to the simpler voltage pulse application shown in Figure 3, here we apply an initial -1 or +1 V pulse, which is followed immediately by a 0.5 V pulse with the opposite polarity. At the same time, a light pulse is introduced with different time delays with respect to the applied voltage profile, shown schematically in the top parts of Figure 4a and b. The weight update is now dominated by the initial ±1 V or the following ±0.5 V pulse, depending on which part of the voltage profile overlaps with the light pulse. In a neuromorphic network employing STDP-learning, this voltage profile would be applied by a firing artificial neuron. If the input precedes the firing of the neuron (positive Δt), there is a causal relationship between the input and the firing, and the synaptic weight will increase. In our implementation, this is due to the overlap between the light pulse and the -1 V pulse. The resulting increase in photocurrent corresponds to the increase in synaptic weight. Conversely, if the input follows the firing of the neuron (negative Δt), signifying an anti-causal relationship, the weight will be decreased. Here, this is brought about by the overlap of the light pulse with the +0.5 V pulse, resulting in a lower photocurrent. In this way, STDP allows associative learning in a neuromorphic network.

Figure 4a demonstrates that the MPOS shows STDP learning for a -1 to +0.5 V pulse and light pulse inputs with different time delays. Similar to the photocurrent changes in Figure 3, we can obtain the inverse weight changes by applying a +1 to -0.5 V pulse in Figure 4b, resulting in anti-STDP learning. The high symmetry of the (anti-)STDP responses of the synapse allows predictable updates of the synaptic weights, which is important for reliable learning in neuromorphic networks.[40] In both cases, large weight updates of up to 2000% are obtained. This large dynamic range allows for easy distinction of different states of the MPOS. Both positive and negative photocurrents can be achieved within each STDP learning rule, depending on the sign of the time delay. Importantly, this shows that the simple STDP learning rule is sufficient to support the unique advantage of switching between inhibitory and excitatory functioning of the synapse.

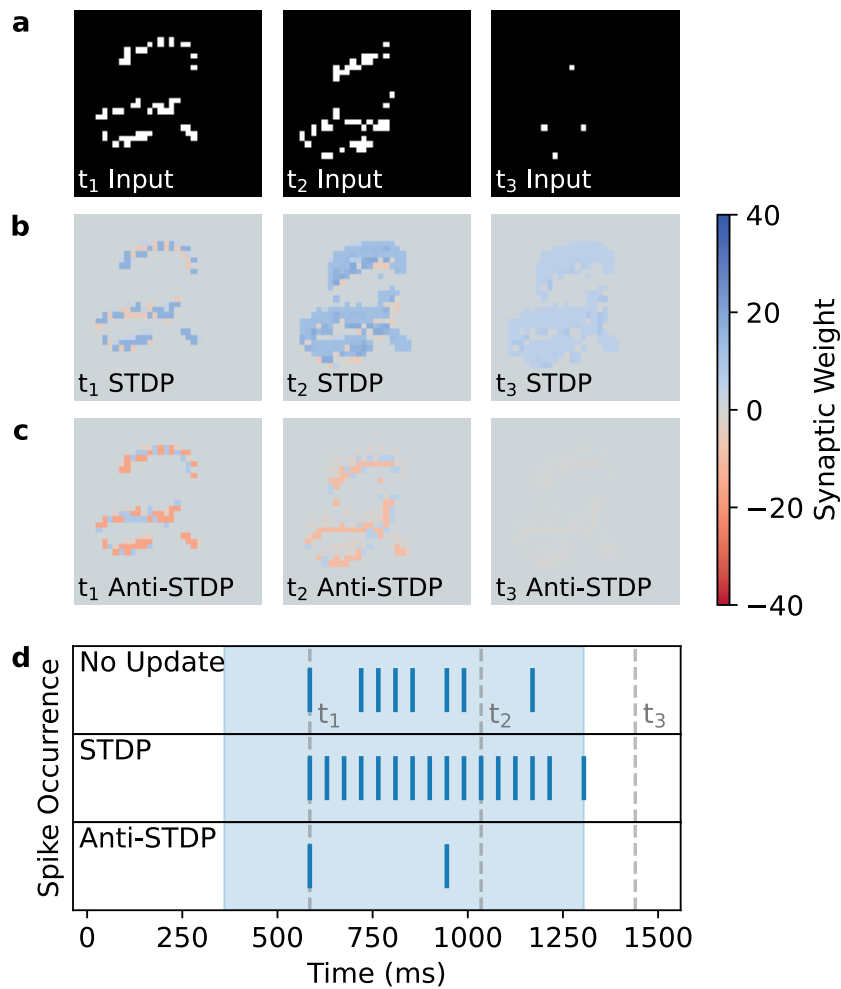

*Figure 5. Simulations of an attention mechanism for arrays of the optoelectronic synapses connected to a simple leaky integrate-and-fire neuron. Frames of an N-MNIST sample number 2 are projected on the arrays over time. **(a)** Frames of the sample at different times of interest. **(b), (c)** The synaptic weights of optoelectronic synapse arrays implementing STDP and anti-STDP learning, respectively, taken after projecting the frames in **(a)** on the arrays. **(d)** Event-plot of the*

*neuron spikes over time for arrays implementing either no synaptic weight updates ("No Update"), or the STDP and anti-STDP weight updates from **(b)** and **(c)**. After each neuron spike, all synaptic weights in the arrays are updated according to their respective update rule. The blue-shaded region indicates the simulation times between 360 ms to 1305 ms where the input number 2 is visible in the N-MNIST frames, which should cause spiking by the neuron. The first neuron spike occurs at time $t_1$ (585 ms), after input of the left panel in **(a)**. The spike causes positive synaptic weight changes for the STDP array (left panel in **(b)**), and negative weights changes for the anti-STDP array (left panel in **(c)**). At time $t_2$ (1035 ms) the frame in the middle panel in **(a)** is projected on the arrays. The resulting photocurrent is not high enough to cause spiking for the array that does not implement weight updates. The higher photocurrents output by the STDP array (middle panel in **(b)**) are large enough to cause spiking, while the negative photocurrents output by the anti-STDP array (middle panel in **(c)**) suppress spiking. After 1305 ms, only noise is projected on the arrays and no further spiking is recorded, resulting in decay of the synaptic weights. The noise projected on the arrays at time $t_3$ (1440 ms) is shown in the right panel in **(a)**. The decaying synaptic weights are shown in the right panels of **(b)** and **(c)**.*

We envision that these results could be particularly interesting for the development of neuromorphic detectors that process visual information. Modern state-of-the-art software neural networks based on transformer models achieve high classification accuracies by employing an attention mechanism to focus only on the relevant regions of an image.[3,4] A detector consisting of dense cross-bar arrays of the MPOS could employ a similar mechanism through STDP. To showcase this application, we simulated two-dimensional arrays of the synapses connected to a leaky integrate-and-fire neuron. The arrays are illuminated with a moving hand-written digit sample from the N-MNIST dataset,[49] and the resulting photocurrent is integrated by the neuron. Neuron spikes are used as a feedback signal to update the synaptic weights depending on the simultaneous illumination condition of each synapse. Feedback neuron spikes were simulated as simple -1 V pulses, as in Figure 2b, or the (anti-)STDP pulses from Figure 4, to represent all experimentally demonstrated update rules in this work (see Supplementary Note 1 for more details). The Supplementary Movies show how the synaptic weights are updated over time by (anti-)STDP or -1 V pulses, or not at all, as a sample of the N-MNIST dataset is projected on the arrays.

Figure 5 shows three points in time of the simulations of arrays implementing STDP and anti-STDP learning. Figure 5a shows the three input frames of the N-MNIST sample at these times. The first neuron spike occurs at time $t_1$, as the left image is presented. The left frame in Figure 5b shows the synaptic weights after an STDP weight update by this spike. The weights of the synapses that were illuminated right before the neuron spiked have increased, in accordance with the measurements in Figure 4a. As is evident from Figure 5d, the increased weights, i.e. higher photocurrents, of the synapses in the array cause a higher spiking frequency of the neuron with subsequent inputs. On the other hand, the left frame of Figure 5c shows the synaptic weights for an array implementing anti-STDP learning. Synapses that were illuminated right before the spike have their weights decreased, as in Figure 4b. Figure 5d shows that this change reduces the neuron spike frequency for later inputs.

The synaptic weights are updated according to the STDP and anti-STDP learning rules from Figure 4 as the handwritten digit moves downwards between $t_1$ and $t_2$, shown in the middle panel in Figure 5a. STDP learning causes the weights to increase dynamically based on the movement of the digit, as shown in the middle panel of Figure 5b, while they are decreased for anti-STDP learning, illustrated by the middle panel of Figure 5c. After the digit is no longer visible and only noise is presented to the array at time $t_3$ in the right panel in Figure 5a, the volatility of the synapses causes the weights to decay, as demonstrated by the right panels in Figure 5b and c. Importantly, this prevents spiking of the neuron due to noise and resets the weights in the arrays for new input features of interest.

The simulations show that, even though voltage pulses are applied to all synapses in the array, the optoelectronic STDP learning rule we implement in Figure 4a only increases the weights of the synapses detecting the feature of interest. This attention mechanism causes the array to adaptively focus on the digit, allowing the neuron to respond more quickly to the input. Anti-STDP learning, on the other hand, can be implemented to reduce attention, forcing the neuron to ignore specific features. This way, the synapse array combines the filtering and working memory applications of volatile synapses. The transient photocurrent enhancement can be seen as a working memory that keeps track of the location of features of interest to filter visual data and focus only on relevant stimuli. The device design allows for flexible tuning of the response parameters. $FAPbBr_3$ and $FAPb(I_{0.5}Br_{0.5})_3$ layers could be incorporated into the arrays, as demonstrated by Figure S8 and Figure S9, for finer control over attention by not only considering the light intensity, but also its wavelength. This is especially relevant for more complex input images that contain different colors.[32] Apart from that, Figure S14 and the Supplementary Movies show a more top-down attention mechanism that can be realized by applying voltage pulses to only a subset of synapses in the array. A similar algorithm has been proposed before, and could be implemented for more complex inputs, for example in image processing for autonomous vehicles.[50]

Recently, similar detectors of volatile memristive devices employing an attention mechanism have been proposed, based on synaptic transistors of two-dimensional materials,[50,51] or a metal oxide active layer.[32] On the device level, the two-terminal architecture and the easier deposition of the halide perovskite layer over large areas make our MPOS easier to scale compared to these implementations. Moreover, light absorption by the halide perovskite layer can be tuned into the visible spectrum by altering its composition, giving more control over the attention mechanism. On the algorithm side, the optoelectronic (anti-)STDP updates we present here remove the need to determine which synapses to update by a top-down approach and instead allow a more easily implemented bottom-up attention mechanism. This way, the synapses leverage the unique combination of light-dependent mixed ionic-electronic conductivity, tunability of the bandgap, and facile deposition of halide perovskites to enable neuromorphic detectors with more biologically plausible learning.

# Conclusion

In summary, we have demonstrated microscale volatile optoelectronic synapses made from $MAPbI_3$, $FAPbBr_3$, and $FAPb(I_{0.5}Br_{0.5})_3$ halide perovskites. The MPOS leverage mobile ions to form a transient electric field after applying a bias voltage, resulting in volatile photocurrent changes upon illumination of the device. We have shown that the magnitude and polarity of the photocurrent are tunable with the applied voltage and light intensity due to the higher iodide-vacancy mobility under illumination and the larger electric field build-up for higher applied voltages. Important features of the MPOS are the separation of electronic writing and photonic read-out, preventing accidental changes in the synaptic weight, and the accessibility of both positive and negative synaptic weights, which allows easy switching between excitatory and inhibitory functioning. The MPOS showed learning based on STDP weight updates, and we simulated how this learning rule could be implemented for a bottom-up attention mechanism in neuromorphic sensors to focus on regions of interest in visual data. Sensors that are only sensitive to parts of the visible spectrum could be fabricated by changing the halide perovskite in the synapse, giving further control over the attention mechanism. The easy scalability of our two-terminal microscale devices and the broad absorption range make the MPOS particularly well-suited for this application. Moreover, the mild fabrication conditions allow easy implementation of the MPOS with existing memristive or CMOS-based neuromorphic networks, extending even to novel implementations on flexible substrates.

# Methods

Si wafers with a 100 nm dry thermal oxide layer were purchased from Siegert Wafer. $PbI_2$ (99.99%), $PbBr_2$ (99.99%), and formamidinium iodide (FAI, 99.99%) were purchased from TCI. Methylammonium iodide (MAI) was purchased from Solaronix. $Al(CH_3)_3$ (97%), formamidinium bromide (FABr, >98.0%) and anhydrous DMF, DMSO, and chlorobenzene were purchased from Sigma-Aldrich. MA-N1410 resist and its corresponding MA-D533/s developer were purchased from Micro Resist. All materials were used without further purification.

### Fabrication of the optoelectronic synapse

Gold bottom electrodes were patterned on the silicon wafer with the thermal oxide layer using a lift-off process with MA-N1410 photoresist. The resist was exposed to UV light in a Süss MA6/BA6 mask aligner. The exposed resist was developed in MA-D533/s. A chrome adhesion layer (5 nm) and the gold electrode layer (80 nm) were deposited on the patterned resist by e-beam physical vapor deposition. Lift-off was performed by soaking in acetone for one hour. A 15 nm $Al_2O_3$ layer was deposited in a home-built atomic-layer deposition setup at 250 °C, using $Al(CH_3)_3$ and $H_2O$ as the precursor gasses. The gold top electrodes were patterned perpendicular to the bottom electrodes using the same UV lithography procedure.
Inside a $N_2$-filled glovebox, $MAPbI_3$, $FAPbBr_3$, and $FAPb(I_{0.5}Br_{0.5})_3$ precursors were mixed by dissolving stoichiometric 1.1 mmolar mixtures of the respective solids in 1 mL DMF and 0.1 mL DMSO. The precursors were filtered with 0.2 μm PTFE filters and spin coated over the gold electrodes at 4000 rpm for 30 seconds with a SCIPRIOS SpinCoating Robot.
Chlorobenzene (250 μL per substrate) was added after 5 seconds of spinning as an antisolvent to induce crystallization. The substrates were annealed at 100 °C for 10 minutes directly after spin coating. The devices were then encapsulated by adding a drop of Blufixx epoxy on the active area. A glass coverslip was dropped on the epoxy, which was cured with a UV torch for 1 minute afterward. The same spin coating procedure was followed to deposit the three perovskite layers on quartz substrates for absorption measurements.

### UV/Vis absorption measurements

Absorption measurements were performed from 250 to 900 nm with a Perkin Elmer Lambda 750 UV/Vis/NIR spectrophotometer inside an integrating sphere. Deuterium and tungsten-halogen lamp light sources and an InGaAs detector were used for the measurements.

## Photocurrent measurements

All electronic measurements were performed with a Keysight B2902A Precision Source/Measure Unit. One channel of the SMU was used to apply voltage pulses to and measure the current of the synapse, while a second channel was used to drive the 450, 520, or 620 nm high power Cree XLamp XP-E LEDs. Irradiances were measured with a Thorlabs PM100D optical power meter with a S120VC sensor.

## Drift-diffusion simulations

Drift-diffusion simulations were carried out with the software package Setfos by Fluxim. The device parameters are listed in Table 2. We simulated the relaxation of the potential and mobile ion density at 0 V after removing the initially applied voltage of 1 V.

Table 2: Simulation parameters used for the drift-diffusion simulations.

| Parameter | Value | Reference |
|---|---|---|
| Thickness insulator (nm) | 15 | |
| Relative permittivity insulator | 9 | |
| Electron affinity insulator (eV) | 2.5 | |
| Band gap insulator (eV) | 5 | |
| Thickness perovskite (nm) | 50 | |
| Relative permittivity perovskite | 24.1 | [52] |
| Electron affinity perovskite (eV) | 3.9 | [53] |
| Band gap perovskite (eV) | 1.6 | [53] |
| Effective density of states perovskite conduction band (1/cm$^3$) | 8x10$^{18}$ | |
| Effective density of states perovskite valence band (1/cm$^3$) | 8x10$^{18}$ | |
| Mobile positive ion density (1/cm$^3$) | 3x10$^{17}$ | |
| Immobile negative ion density (1/cm$^3$) | 3x10$^{17}$ | |
| Work function electrodes (eV) | 5.1 | |

# Acknowledgements

The work of J.J.B., A.O.A., M.C.S., and B.E. received funding from the European Research Council (ERC) under the European Union's Horizon 2020 research and innovation programme under grant agreement No. 947221. The work is part of the Dutch Research Council NWO and was performed at the research institute AMOLF. The authors thank Marc Duursma, Bob Drent, Igor Hoogsteder, Arthur Karsten, and Laura Juškėnaitė for technical support.

Supporting Information for

# Microscale optoelectronic synapses with switchable photocurrent from halide perovskite

Jeroen J. de Boer[1], Agustin O. Alvarez[1], Moritz C. Schmidt[1], Dimitris Sitaridis[1], Bruno Ehrler[1,*]

[1]LMPV-Sustainable Energy Department, AMOLF, 1098 XG, Amsterdam, the Netherlands
*Correspondence: b.ehrler@amolf.nl

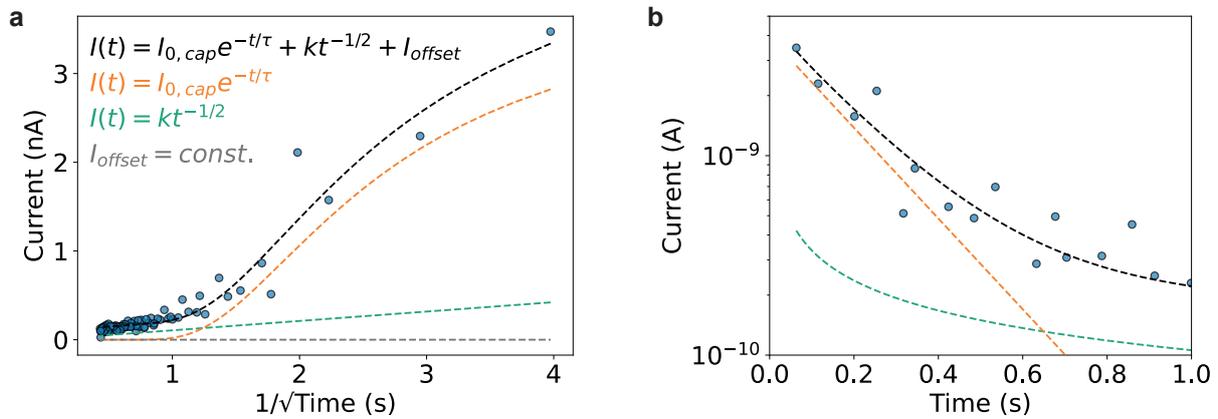

Figure S1. Plots of the measured photocurrents from Figure 2b on different scales for fitting of the data. **(a)** Plot on the $t^{-1/2}$ scale. The linear increase of the photocurrent until approximately 1.5 sec$^{-1/2}$ indicates a current decay with $t^{1/2}$ proportionality for $t > \frac{1}{1.5^2} \approx 0.44$ seconds, in line with a diffusion-limited process. **(b)** Plot on the semi-log scale. The plot shows a linear decrease in the current for the times before approximately 0.5 seconds, which can be explained by a capacitive drift current.

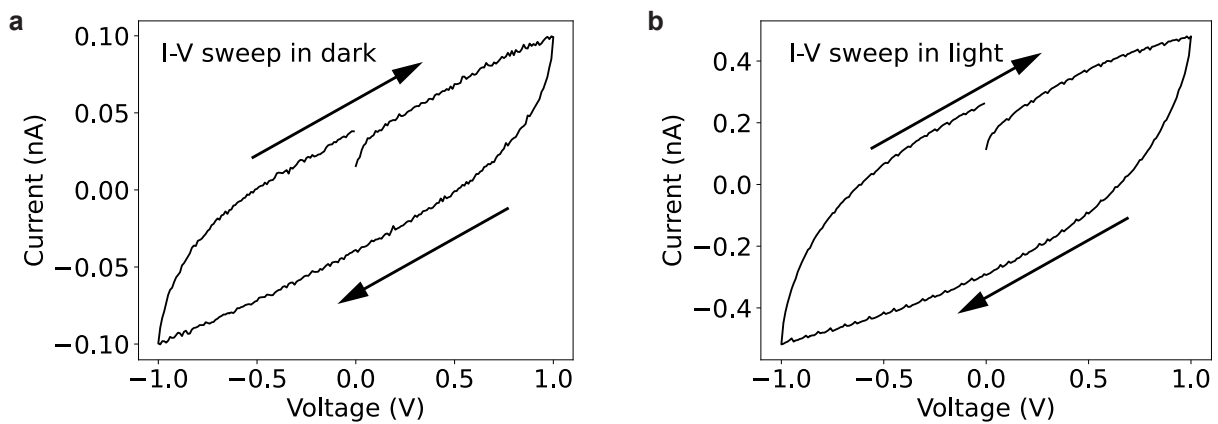

Figure S2. I-V sweeps of the device under different illumination conditions. **(a)** An I-V sweep in the dark. **(b)** An I-V sweep on the same device under illumination with a 520 nm LED, at an irradiance of 0.2 mW/cm$^2$. For both plots, the arrows indicate the current measured at each voltage sweep direction. The plots were obtained by averaging the current over three consecutive measurements to reduce noise. The current responses are typical for a capacitive displacement, combined with a resistive current. A larger current is measured for the device as it is illuminated, indicating a lower resistance. No significant changes in the resistance of the device were measured within either scan.

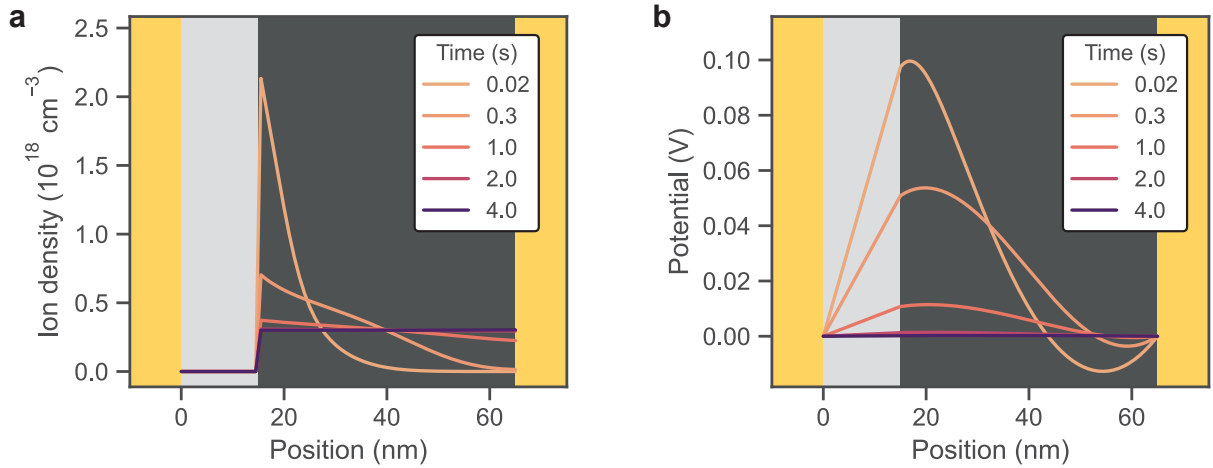

*Figure S3. Drift-diffusion simulation of the device after removal of a -1 V pulse at time = 0. **(a)** The halide vacancy distribution after the pulse. **(b)** The resulting potential in the device after the pulse. The initial accumulation of halide vacancies at the $Al_2O_3$-covered cathode results in a potential in the device. The vacancies redistribute over approximately 4.0 seconds, causing a decay of the potential.*

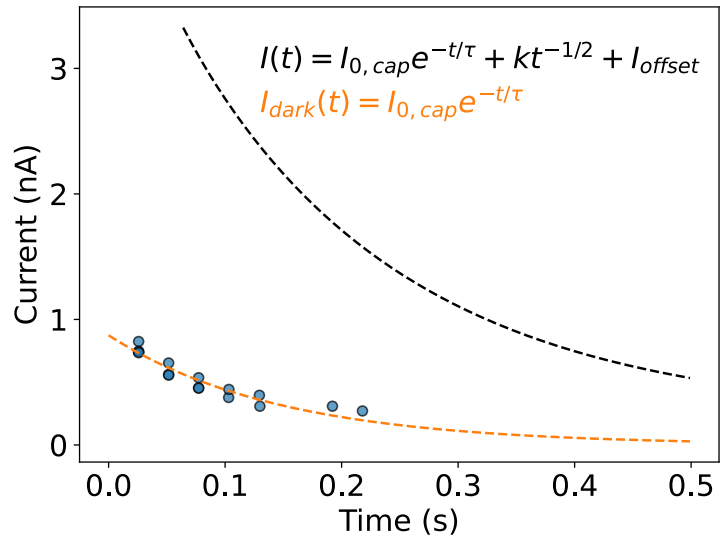

*Figure S4. Fit of the current in dark measured after the -1 V pulse and before the first light pulse in Figure 2a and b in the main text. Data taken from the same 5 measurements as in Figure 2b in the main text. The dark current is fit to an exponential decay and compared to the fit to the measured photocurrents from Figure 2b.*

Figure S4 shows a fit to the current in dark after a -1 V pulse is applied and before application of a light pulse. The data is taken from the same measurements as Figure 2b in the main text. The data is fit with an exponential decay (fitting parameters were $I_{0,cap} = 0.87 \pm 0.04$ nA, and $\tau = 0.15 \pm 0.01$ s). Fitting with an additional diffusion term was not successful, which is likely due to the relatively low contribution of this term for these short timescales. The decay time on the hundreds of milliseconds timescale implies that this is the ionic drift current in the dark. The black curve represents the fit to the photocurrent from Figure 2b. From the fits it follows that the ionic current contributes 17% to the total current at 0.06 seconds, when the first light pulse is applied. For later times, the fits suggest that the relative contribution of the ionic current decreases due to the lack of a diffusion and constant offset term for the current in dark.

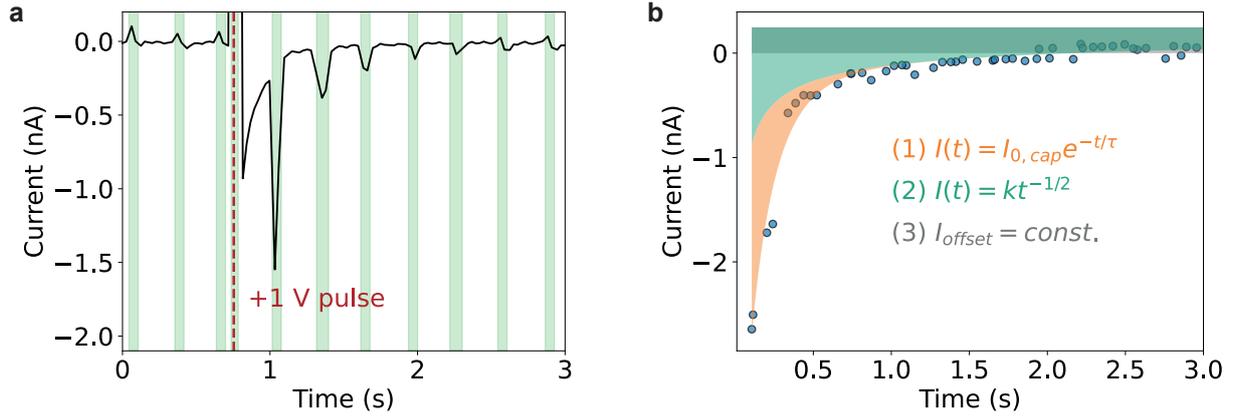

*Figure S5. Setting and reading out of the state of the optoelectronic synapse with a +1 V applied pulse. **(a)** Measured photocurrents over time. Similar to the measurement in Figure 2a, a small initial photocurrent is read out when the device is illuminated with a green LED, indicated by the green regions. After applying the +1 V pulse, indicated by the red dotted line, a negative photocurrent is read out with consecutive light pulses, which decays to the initial photocurrent from before the voltage pulse over time. **(b)** Fitting of the photocurrents in (a) over time, after applying a +1 V at time = 0. The blue markers indicate the measured photocurrents over five measurements. The same fitting equation that considers a combined ionic drift and diffusion process for the current decay, combined with a constant offset current was used to fit the transient photocurrent: $I(t) = I_{0,cap}e^{-t/\tau} + kt^{-1/2} + I_{offset}$.*

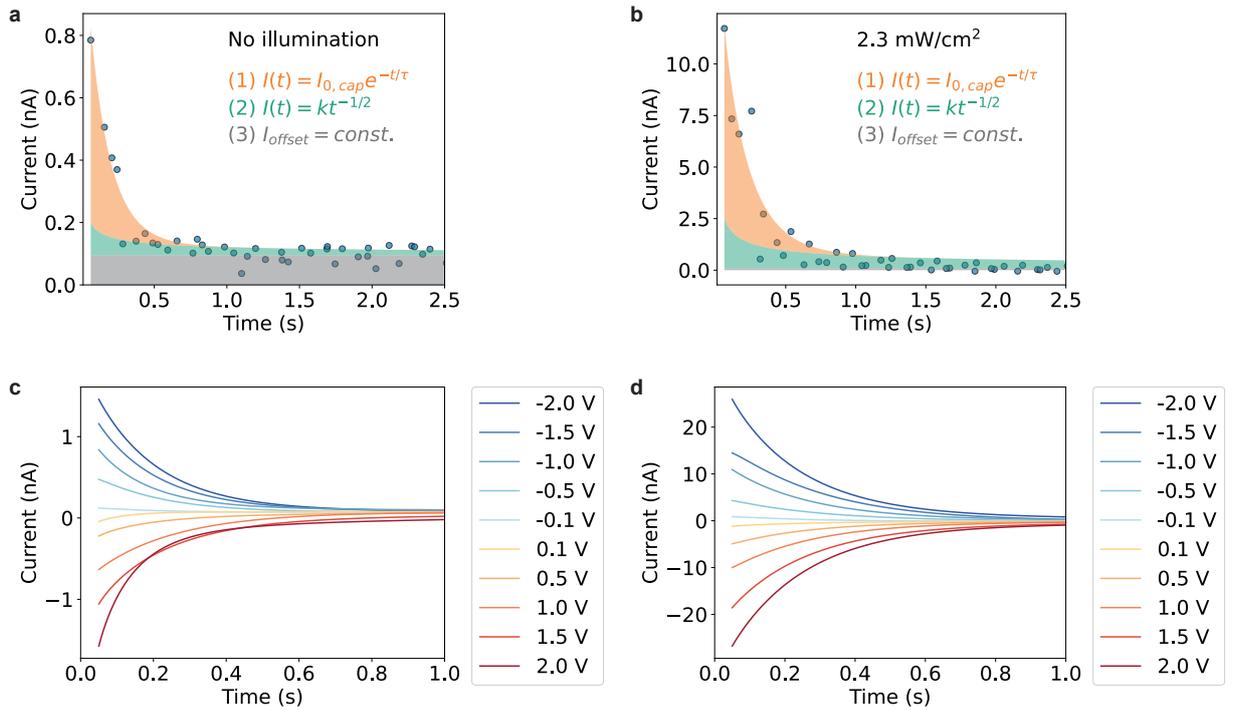

*Figure S6. Fitting of the photocurrents over time for different illumination intensity conditions. **(a)** Shows the photocurrents over time after a -1 V pulse without illumination, while **(b)** shows the data for simultaneous illumination with a 2.3 mW/cm² irradiance. The blue markers indicate the measured photocurrents over five measurements. The transient photocurrent was fit using the same equation as in Figure 2b: $I(t) = I_{0,cap}e^{-t/\tau} + kt^{-1/2} + I_{offset}$. A similar initial drift followed by a diffusion-limited photocurrent decay is obtained for both conditions. Fitting parameters are given in Table S1. **(c)** and **(d)** show fits of transient photocurrent measurements with applied voltage pulses ranging from -2.0 to +2.0 V for no illumination or illumination with a 2.3 mW/cm² irradiance during the voltage pulse, respectively. Larger photocurrent changes are measured for larger voltage amplitudes and for higher irradiance during the voltage pulse.*

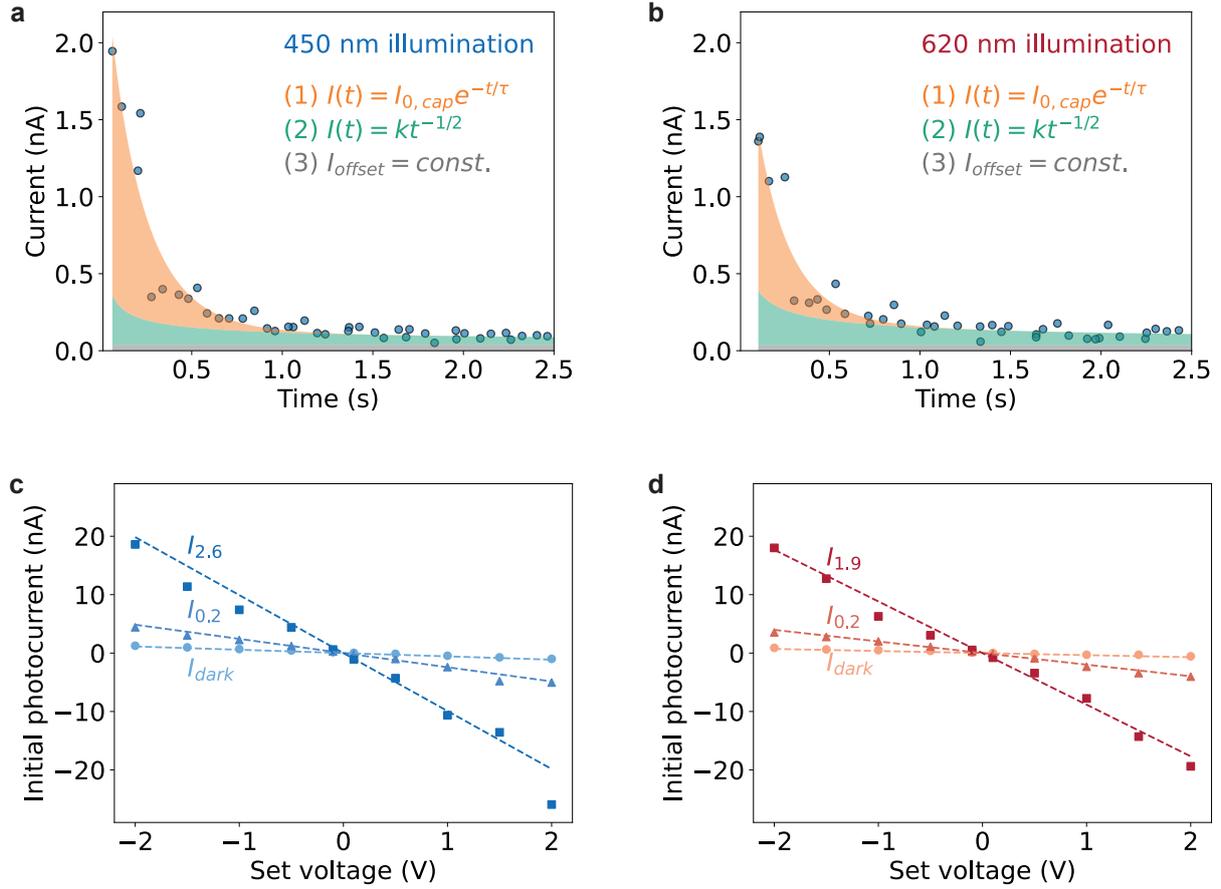

*Figure S7. Fitting of the photocurrents over time for different wavelengths of light. **(a)** Fits of the measured photocurrent after applying a -1 V at time = 0 with simultaneous illumination with 450 nm light, with a 0.24 mW/cm² irradiance. **(b)** The same experiment as in **(a)**, repeated with 620 nm light, with a 0.18 mW/cm² irradiance. The power densities were adjusted to ensure equal photon flux for each of the three wavelengths of light. The blue markers indicate the measured photocurrents over five measurements. Similar to before, the transient photocurrent was fit with: $I(t) = I_{0,cap}e^{-t/\tau} + kt^{-1/2} + I_{offset}$. A similar initial drift followed by a diffusion-limited photocurrent decay is obtained for both wavelengths. Fitting parameters are given in Table S1. **(c)** Comparison of the photocurrents at 0.05 seconds, obtained from the transient current fits, for different applied voltages, and 450 nm light intensities of 2.6 mW/cm² ($I_{2.6}$, squares), 0.2 mW/cm² ($I_{0.2}$, triangles), and no light ($I_{dark}$, circles) during the voltage pulse. Slopes of the linear fits of the initial photocurrents are -9.9, -2.4, and -0.6 nA/V for $I_{2.6}$, $I_{0.2}$, and $I_{dark}$, respectively. **(d)** The same measurements repeated for 620 nm light excitation, with irradiances of 1.9 mW/cm² ($I_{1.9}$, squares), 0.2 mW/cm² ($I_{0.2}$, triangles), and no light ($I_{dark}$, circles) during the voltage pulse. The obtained slopes of $I_{1.9}$, $I_{0.2}$, and $I_{dark}$ are respectively -8.8, -2.0, and -0.4 nA/V.*

Figure S7 shows similar photocurrent changes observed in Figure 2b and 3b for 450 and 620 nm light. For both light sources, the photocurrent is enhanced after applying a -1 V pulse, as shown in Figure S7a and b. After initial drift-dominated decay, the photocurrent then decays further by a diffusion-limited process. Fitting parameters of the measurements are given in Table S1.

Similar to the 520 nm experiments in Figure 3b, the photocurrent changes for the 450 and 620 nm light sources are linear with respect to the input voltage and more significant under higher irradiances, as follows from Figure S7c and d and the fitting parameters. The larger photocurrent changes for the 450 nm light can be explained by the higher absorption by the $MAPbI_3$ layer for this wavelength of light (see the UV/Vis absorption spectrum in Figure 1c).

*Table S1. Fitting parameters for the measurements performed with different irradiance (Figure S6a and b) or wavelength (Figure S7a and b) during the -1 V pulse. Errors indicate one standard deviation.*

| Illumination conditions | $I_{0,cap}$ $(nA)$ | $\tau$ $(s)$ | $k$ $(nA\sqrt{s})$ | $I_{offset}$ $(nA)$ |
|---|---|---|---|---|
| No illumination | $0.97 \pm 0.09$ | $0.14 \pm 0.01$ | $0.026 \pm 0.016$ | $0.074 \pm 0.011$ |
| 520 nm, 2.3 mW/cm² | $13.0 \pm 1.4$ | $0.19 \pm 0.02$ | $0.61 \pm 0.33$ | $-0.33 \pm 0.22$ |
| 450 nm, 0.24 mW/cm² | $2.3 \pm 0.2$ | $0.20 \pm 0.02$ | $0.080 \pm 0.054$ | $0.038 \pm 0.035$ |
| 620 nm, 0.18 mW/cm² | $1.9 \pm 0.2$ | $0.18 \pm 0.02$ | $0.12 \pm 0.04$ | $0.035 \pm 0.028$ |

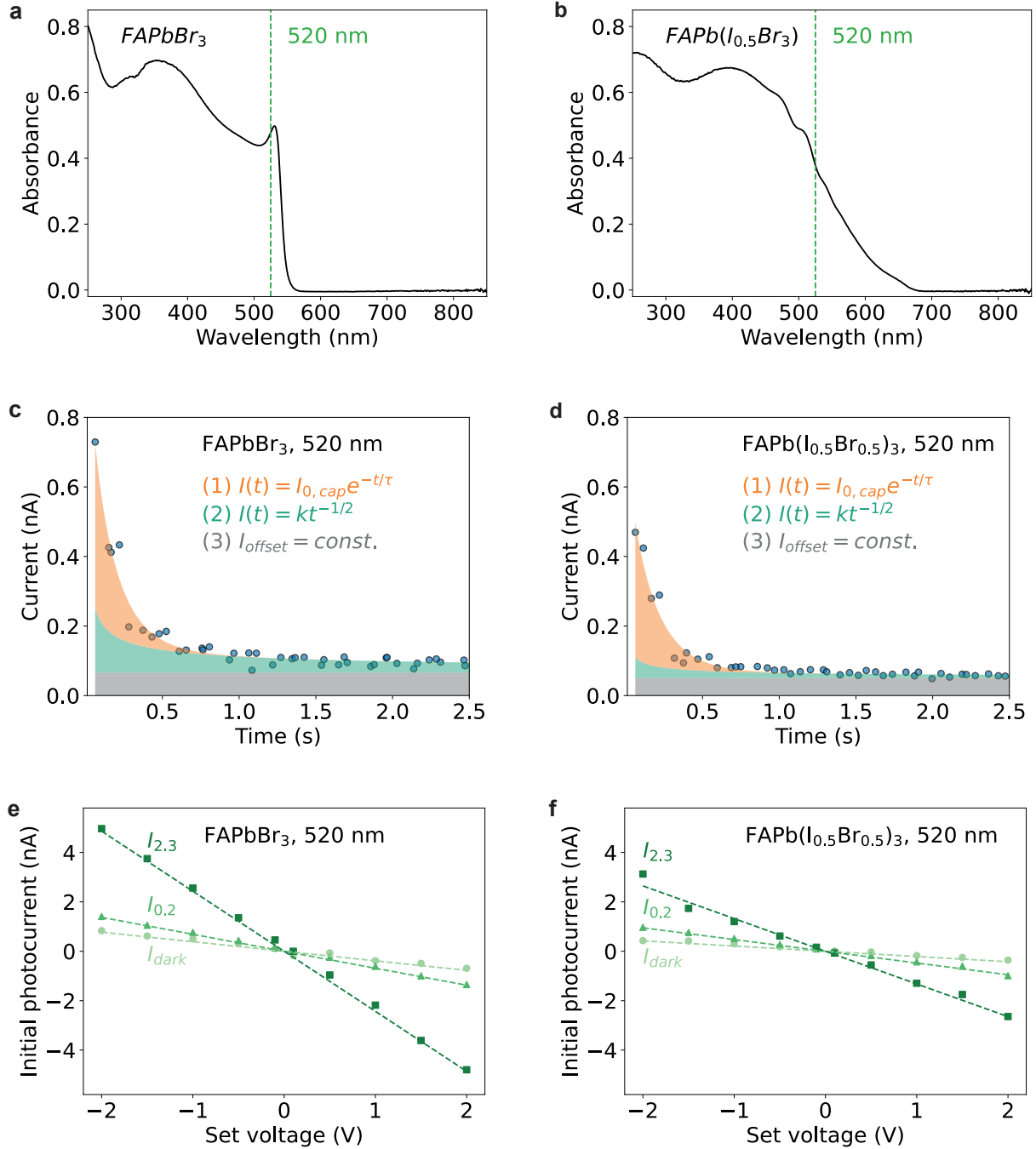

*Figure S8. Optoelectronic synapse measurements on devices with FAPbBr$_3$ and FAPb(I$_{0.5}$Br$_{0.5}$)$_3$ halide perovskite active layers with 520 nm light. **(a), (b)** Absorption spectra of, respectively, FAPbBr$_3$ and FAPb(I$_{0.5}$Br$_{0.5}$)$_3$ on quartz substrates. The absorption onsets shift to lower wavelengths for perovskite layers with a larger bromide content. **(c), (d)** Fits to transient photocurrent measurements with a -1 V pulse at t = 0 seconds and 520 nm light illumination with a 0.2 mW/cm$^2$ irradiance. Both devices show similar decays, with a larger photocurrent measured for the FAPbBr$_3$-device. **(e), (f)** Linear fits to initial photocurrents obtained for different set voltages and illumination intensities of 2.3 mW/cm$^2$ (I$_{2.3}$, squares), 0.2 mW/cm$^2$ (I$_{0.2}$, triangles), or no illumination (I$_{dark}$, circles). For the FAPbBr$_3$ device, the slopes of I$_{2.3}$, I$_{0.2}$, and I$_{dark}$ were -2.4, -0.7, and -0.4 nA/V, respectively. The slopes of I$_{2.3}$, I$_{0.2}$, and I$_{dark}$ for the FAPb(I$_{0.5}$Br$_{0.5}$)$_3$ device were respectively -1.3, -0.5, and -0.2 nA/V.*

Figure S8 shows that optoelectronic synapses can be fabricated with FAPbBr$_3$ and FAPb(I$_{0.5}$Br$_{0.5}$)$_3$ active layers as well. The absorption spectra of FAPbBr$_3$ and FAPb(I$_{0.5}$Br$_{0.5}$)$_3$ in Figure S8a and b, respectively, show that the absorption onset of these perovskites shifts to shorter wavelengths for more bromide-containing perovskites.

Figure S8c and d show measurements of photocurrent modulation with a -1 V pulse. Both perovskites show similar photocurrent enhancements and decays, again first dominated by drift, followed by diffusion at later times. Figure S8e and f show the expected higher photocurrent enhancement for larger voltage amplitudes and illumination intensities. The slightly larger photocurrent enhancement of the $FAPbBr_3$ device can be explained by the higher 520 nm light absorption of this film.

Figure S9a and b show similar photocurrent decays for measurements with 450 nm illumination. Both devices show slightly higher photocurrents compared to the 520 nm illumination conditions in Figure S8c and d. The fits to the initial photocurrents with respect to the set voltage in Figure S9c and d are also slightly steeper compared to those in Figure S8e and f. Both discrepancies can be explained by the higher absorbance for shorter wavelengths of both perovskites, as follows from the absorption spectra in Figure S8a and b. Fitting parameters for the measurements in Figure S8c and d and Figure S9a and b are given in Table S2. For both wavelengths, the photocurrent enhancement is much less significant compared to the measurements on the $MAPbI_3$ device in Figure 2b and Figure 3b. This can be explained by the higher mobility of iodide vacancies in $MAPbI_3$ perovskites.[1,2]

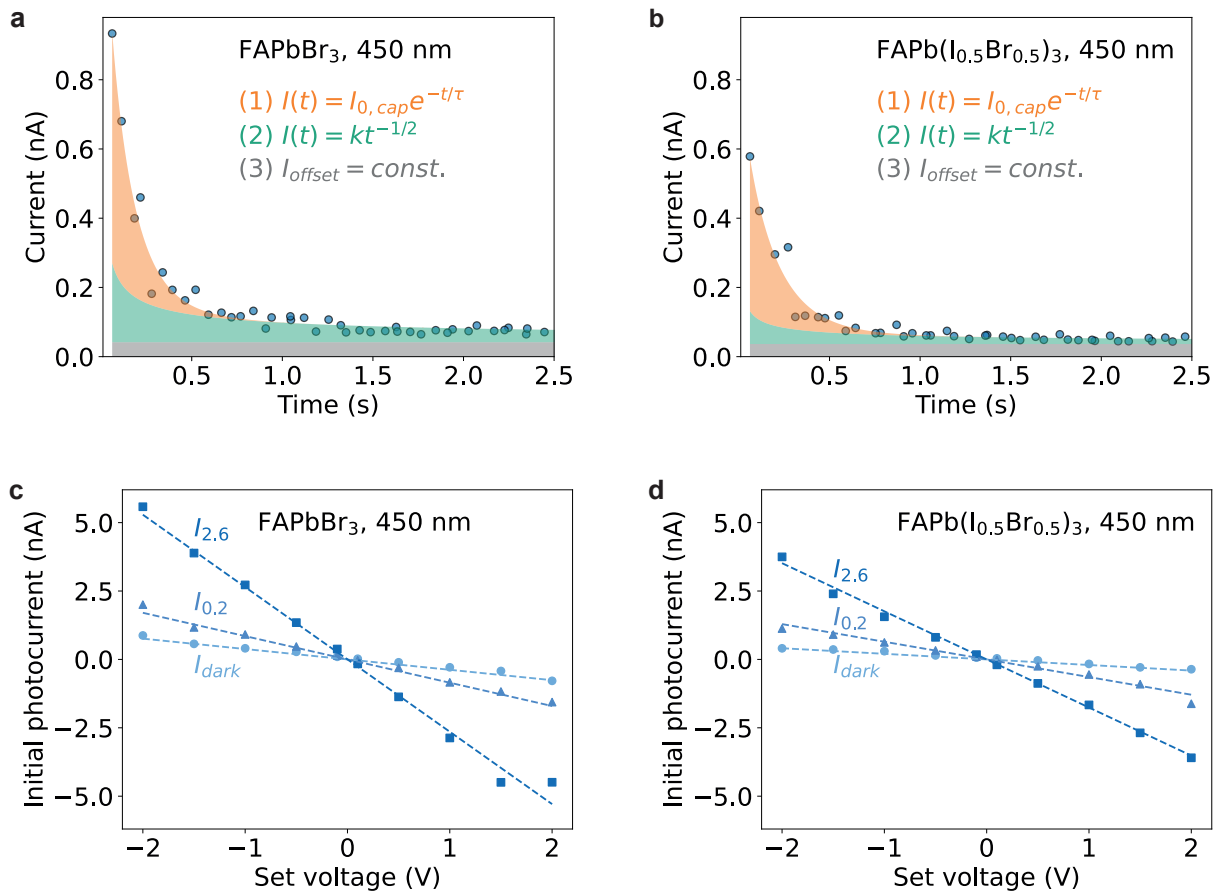

Figure S9. Optoelectronic synapse measurements on devices with $FAPbBr_3$ and $FAPb(I_{0.5}Br_{0.5})_3$ halide perovskite active layers with 450 nm light. **(a), (b)** Fits to transient photocurrent measurements with a -1 V pulse at t = 0 seconds and 450 nm light illumination with a 0.2 mW/cm² irradiance. As in Figure S8, photocurrent decays are similar for both devices, with a larger initial photocurrent measured for the $FAPbBr_3$ device. **(c), (d)** Linear fits to initial photocurrents obtained for different set voltages and illumination intensities. Irradiances were 2.6 mW/cm² ($I_{2.6}$, squares), 0.2 mW/cm² ($I_{0.2}$, triangles), or the devices were kept in dark during the voltage pulse ($I_{dark}$, circles). For the $FAPbBr_3$ device, slopes of -2.6, -0.9, and -0.4 nA/V were found for $I_{2.6}$, $I_{0.2}$, and $I_{dark}$, respectively. For the $FAPb(I_{0.5}Br_{0.5})_3$ device, the slopes of $I_{2.6}$, $I_{0.2}$, and $I_{dark}$ were -1.8, -0.6, and -0.2 nA/V, respectively.

Table S2. Fitting parameters for the measurements on the FAPbBr$_3$ and FAPb(I$_{0.5}$Br$_{0.5}$)$_3$ devices in Figure S8c and d and Figure S9a and b. Errors indicate one standard deviation.

| Device | $I_{0,cap}$ (nA) | $\tau$ (s) | $k$ (nA$\sqrt{s}$) | $I_{offset}$ (nA) |
|---|---|---|---|---|
| FAPbBr$_3$ (520 nm) | 0.71 ± 0.05 | 0.16 ± 0.01 | 0.046 ± 0.010 | 0.066 ± 0.006 |
| FAPbBr$_3$ (450 nm) | 1.0 ± 0.1 | 0.14 ± 0.01 | 0.056 ± 0.01 | 0.042 ± 0.07 |
| FAPb(I$_{0.5}$Br$_{0.5}$)$_3$ (520 nm) | 0.56 ± 0.04 | 0.17 ± 0.01 | 0.015 ± 0.008 | 0.050 ± 0.005 |
| FAPb(I$_{0.5}$Br$_{0.5}$)$_3$ (450 nm) | 0.63 ± 0.04 | 0.18 ± 0.01 | 0.023 ± 0.009 | 0.037 ± 0.006 |

## Supplementary Note 1. Simulating optoelectronic synapse arrays implementing different learning rules

The simulations in Figure 5, Figure S14, and the Supplementary Videos were performed with a custom Python module. The module used NumPy (version 1.26.4) for numerical operations, run in Python version 3.11. The simulations are illustrated schematically in Figure S10a and b. A 34-by-34-pixel sample of the N-MNIST test set was binarized and binned into 2.5 ms frames. Binarized input frames are represented by the two-dimensional matrix $X(t)$, where each element is either 1 (illuminated pixels), or 0 (dark pixels), so that $X(t) \in \{0,1\}^{34\times 34}$. Illumination was modeled based on the 520 nm, 0.2 mW/cm$^2$ data in the main text. Each frame is sequentially projected onto the 34-by-34 volatile synapse array $W(t)$, where each pixel $X_{mn}(t)$ is used as input for its corresponding synapse $W_{mn}(t)$. The timestep between frames is increased from 2.5 to 45 ms to match the timesteps in the experimental measurements. This leads to an artificial "slowing" of the video data. The 45 ms timestep is a limitation of our experimental setup, and could be overcome with a faster electrical characterization of the MPOS.

For each timestep, first the total photocurrent output of the synapse array is calculated based on $X(t)$ and $W(t)$, as shown in Figure S10a. Figure S10b shows an example calculation of the output current. This current is then used to update the membrane potential of a leaky integrate-and-fire (LIF) neuron, which fires a spike if the membrane potential reaches a threshold. The spike is applied to the synapses in the array as a feedback signal. In actual implementations, the spikes can also be propagated to following layers of a more complex network. The adaptive focusing on features of interest could help these networks with, for example, classification tasks.[3] Finally, the synaptic weights in the array are updated. If a feedback spike is provided, the synaptic weights are updated based on the voltage profile of the spike and the simultaneous illumination conditions. If no feedback spike is provided, the synaptic weights decay to a steady-state value. This Supplementary Note first briefly explains calculations of the photocurrent output by the synapse array and the LIF neuron membrane potential updates. Next, it describes how the synaptic weights are obtained for different feedback spikes.

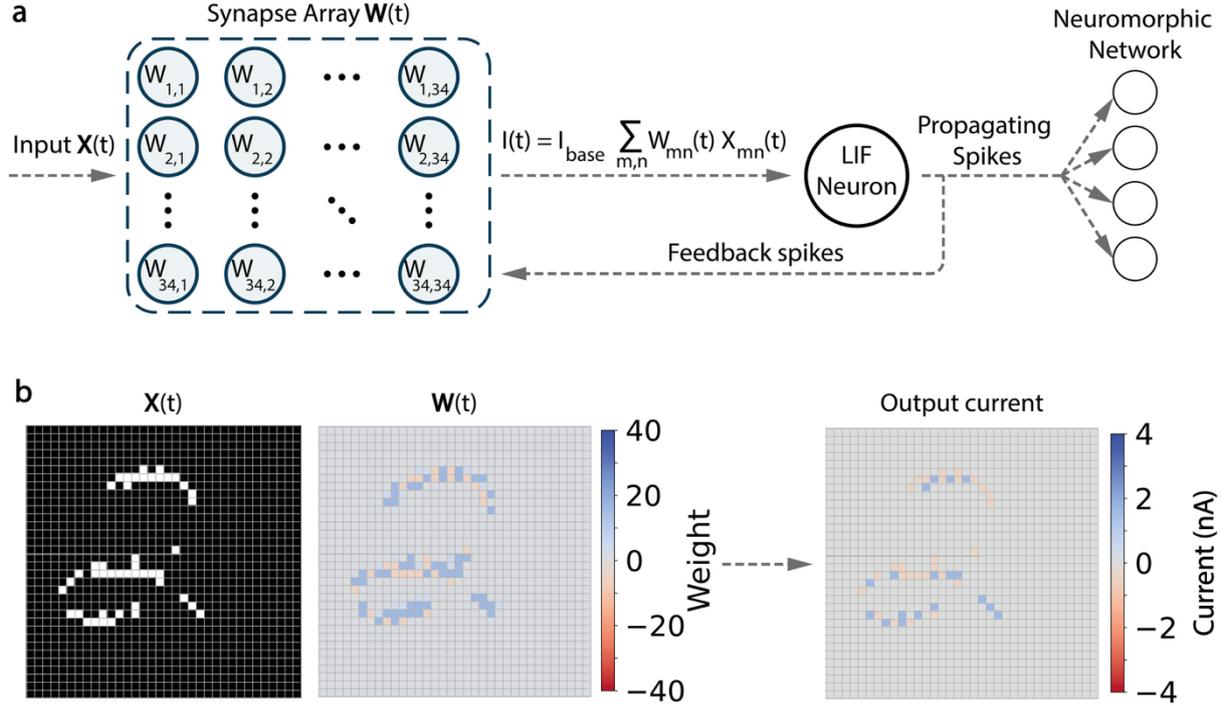

*Figure S10. Schematic representation of the simulations. **(a)** Input frames **X**(t) are projected on an array of synapses **W**(t). Each synapse outputs a weight-dependent current if it is illuminated. All currents are summed and integrated by a leaky integrate-and-fire neuron. Spike outputs by the neuron are used as feedback to update the synaptic weights. The spikes can also be propagated to a neuromorphic network for further processing of the input. **(b)** Example calculation of the current for the input frame **X**(t) and synapse array **W**(t) after the STDP weight update at $t_1$ in Figure 5b. The currents of all pixels are summed to obtain I(t).*

The photocurrent generated by each synapse is determined by multiplying its binary input, $X_{mn}(t)$ with a base current $I_{base}$ of 95 pA (the constant offset photocurrent $I_{offset}$ in Figure 2b of the main text), and a scale factor (synaptic weight), $W_{mn}(t)$. All currents are summed, giving a total current output of the synapse array of:

$$I(t) = I_{base} \cdot \langle \mathbf{W}(t), \mathbf{X}(t) \rangle = I_{base} \cdot \sum_{m=1}^{34} \sum_{n=1}^{34} W_{mn}(t) X_{mn}(t) \qquad (1)$$

The membrane potential of the LIF neuron is updated after each frame based on this current. We set the threshold (1 V), characteristic time (50 ms), and the resistance (2.3 × $10^8$ Ω) of the neuron to obtain appropriate spike rates for the given inputs. After the threshold voltage is reached, the neuron outputs a spike, and the membrane potential of the neuron is reset to 0 V.

The spike applied to the synapse array causes an update of the synaptic weights. The volatility of the synapses is modeled as a weight decay to a steady-state value of 1.0 in timesteps where no feedback spike is applied. All weight decays are modeled based on an exponential $I_{0,cap} e^{-t/\tau}$ drift term. From the fit in Figure 2b in the main text, it follows that the photocurrent decays predominantly by this term. The decay by the $kt^{-1/2}$ diffusion term is only minor and is therefore ignored in the simulations for simplicity. This approximation allows us to describe the synaptic weight changes based on the charging and discharging of a capacitor, where the capacitor voltage is due to ion accumulation at the cathode (per the left panel in Figure 2c).[4] The photocurrent response is determined by this voltage, which is therefore used as a measure of the synaptic weight. With the capacitor approximation, the change in voltage in the device for simple voltage pulses, as in Figure 3 can be described as an RC step response:

$$V(t) = V_{sup} + (V_{init} - V_{sup})e^{-t/\tau} \tag{2}$$

where $V(t)$ is the time-dependent voltage in the synapse induced by ion accumulation, $V_{sup}$ the spike voltage applied to the device, $V_{init}$ the voltage in the device before the update pulse, $t$ is the duration of the pulse, and $\tau$ is the characteristic time. As the timestep in the simulations is constant, equation 2 can be rewritten as a first-order linear recurrence relation:

$$V_{i+1} = V_{sup} + (V_i - V_{sup})e^{-\frac{t_1}{\tau}} = V_i e^{-\frac{t_1}{\tau}} + a \tag{3}$$

where $a = V_{sup}(1 - e^{-\frac{t_1}{\tau}})$, $V_i$ and $V_{i+1}$ are the voltage in the device at timestep $i$ and $i + 1$, respectively. In our simulations, $t_1$ is the 45 ms timestep. In the absence of a feedback spike ($V_{sup} = 0$), $a = 0$ and the voltage decays exponentially. The photocurrent is assumed to be directly proportional to this voltage, which follows from the linear increases in photocurrent with applied voltage magnitude in Figure 3b. This assumption is further supported by the constant resistance observed in the I-V sweeps in light and dark in Figure S2.

From the measurements in Figure 2b and Figure S6a, it follows that a higher photocurrent is extracted after a −1 V pulse is applied while the device is illuminated compared to if the device is kept in dark during the −1 V pulse. This can be captured by equation 3 as a difference in the characteristic time τ during the application of the feedback spike voltage. Mechanistically, the difference in τ can be explained by a higher ionic conductivity under illumination.[5] The characteristic time if the device is in dark can be approximated as $\tau_{dark} = 190$ ms based on the fit in Figure 2b, where the device is kept in dark the majority of the time. Assuming a linear relation between the photocurrent and the built-in voltage, $\tau_{light}$ can be found by setting:

$$\frac{I_{0,light}}{I_{0,dark}} = \frac{V_{i+1,light}}{V_{i+1,dark}} \tag{4}$$

Where $I_{0,light}$ and $I_{0,dark}$ are obtained from the drift-term fit in Figure 2b and Figure S6a, respectively, and $V_{i+1,light}$ and $V_{i+1,dark}$ are obtained from equation 3, setting $\tau = \tau_{dark}$ for $V_{i+1,dark}$, and $V_i = 0$ V for both voltages, per the experimental conditions. From equation 4 we obtain $\tau_{light} = 24$ ms.

Changes to the induced electric field in the device $V_i$ for different illumination conditions and feedback pulse voltages can now be described using $\tau_{dark}$ and $\tau_{light}$ and equation 3. The synaptic weight $W_i$ is a scale factor of the base current that relates this $V_i$ to the output photocurrent $I_i$:

$$I_i = I_{base} \cdot W_i(V_i) \tag{5}$$

Before any voltage is applied to the device, the device is in steady-state conditions. Consequently, the device outputs only the base current upon illumination, for which we define the synaptic weight as 1.0:

$$W_i(V_i = 0) \equiv 1.0 \tag{6}$$

From the linear dependence of the photocurrent $I_i$ on the induced electric field, $V_i$, and of the photocurrent on the synaptic weight in equation 5, it follows that the synaptic weight should also depend linearly on $V_i$. Also considering equation 6, the synaptic weight can be calculated as:

$$W_i(V_i) = 1 + \alpha V_i \tag{7}$$

where $\alpha$ is a constant scale factor. To determine $\alpha$, we calculate $W_{i+1}(V_{i+1})$ and $V_{i+1}$ for the measurement in Figure 2b using equations 5 and 3, respectively. From the obtained values ($W_{i+1}(V_{i+1}) = 31.6$ and $V_{i+1} = -0.85$ V), we calculate $\alpha = -\frac{30.6}{0.85 \text{ V}}$. Substituting into equation 7 yields:

$$W_i(V_i) = 1 - \frac{V_i}{0.85 \text{ V}} \times 30.6 \tag{8}$$

Finally, a first-order linear recurrence relation of $W_{i+1}$ in terms of $W_i$ can be found by substituting equation 3 into equation 8:

$$W_{i+1} = e^{-\frac{t_1}{\tau}} W_i + b \tag{9}$$

with $b = 1 - e^{-\frac{t_1}{\tau}} - \frac{a}{0.85 \text{ V}} \times 30.6$, and $a$ the term of equation 3.

To validate the derived expressions for the synaptic weights and corresponding photocurrents, we used equation 5 and 9 to reproduce experimental results in the main text. Figure S11a compares the calculated weight and photocurrent for different applied voltages $V_{sup}$ to the experimental results from Figure 3b. Figure S11b compares the weight and photocurrent decay obtained from the simulations with the results from Figure 2b. Both are described well by the simulations. The simulated weight changes slightly underestimate the values obtained from experiment because of the lack of the diffusion component.

As an illustration, Figure S11c shows a simple simulation of weight updates of a synapse when a random voltage and light profile is applied. Similar to the plot in Figure S11a, larger weight updates are found for voltage pulses that overlap with light pulses. This simple simulation shows how the weight increases logarithmically with successive applied pulses, and decays exponentially when no voltage is applied, as expected from the approximation of the weight changes as charging and discharging of a capacitor.

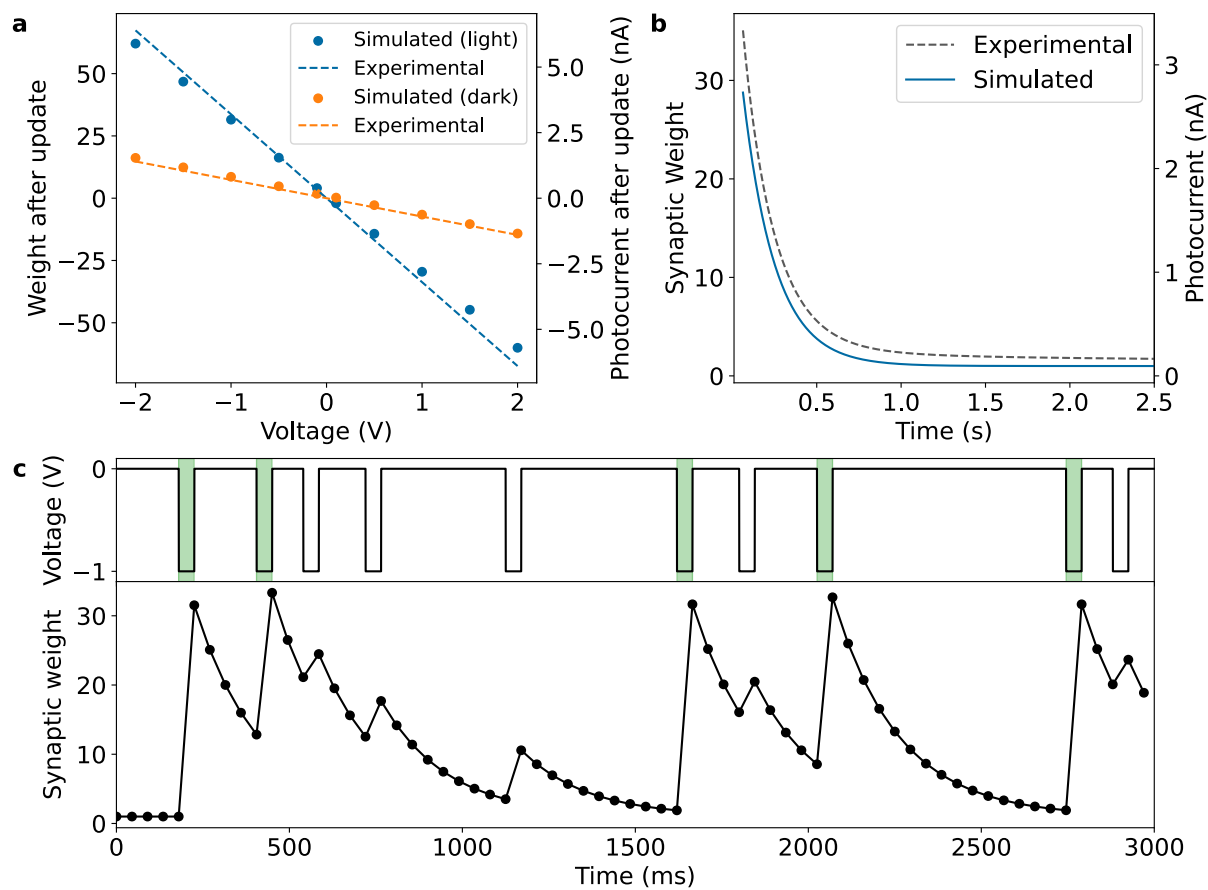

*Figure S11. Comparison of simulations of weight updates of a single synapse to experiments to validate the simulation results.* **(a)** *Comparison of weight changes with respect to voltage with experimental results.* **(b)** *Comparison of the simulated weight decay over time with experimental results. In both* **(a)** *and* **(b),** *the simulated weights slightly underestimate the synaptic weights.* **(c)** *Simulation of synaptic weight updates based on a random voltage and light profile. The random voltage profile is shown in the top panel. Green shaded regions indicate simultaneous illumination of the device. The synaptic weight in the bottom panel increases when a voltage is applied. Every marker represents the synaptic weight at a 45 ms timestep. Increases are more significant if the device is illuminated during a -1 V pulse.*

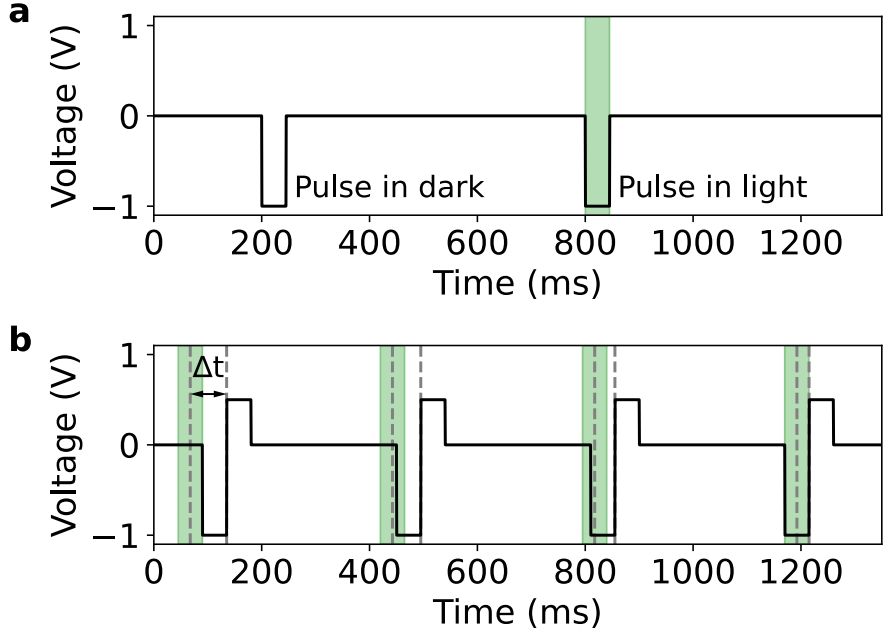

*Figure S12. Schematic of different illumination conditions during applied voltages that are considered by the simulations. (a) Simple voltage pulses (-1 V in this plot) are either applied in the dark, or with simultaneous illumination, shown by the green shaded area. (b) -1 V to +0.5 V STDP pulses overlap partially with the light pulses. The overlap is expressed as the time difference of the center of the STDP voltage profile and the center of the 45 ms light pulse, Δt. The schematic shows examples where Δt is positive and decreasing with each pulse.*

Simulations implementing (anti-)STDP feedback spikes from Figure 4 follow a similar procedure. First, the total photocurrent is determined using equation 1, after which the membrane potential of the same LIF neuron is updated and compared to its threshold. Finally, the synaptic weights are updated. Weight decay was simulated based on equation 3. However, determining $V_i$ for the more complex $\mp 1$ V to $\pm 0.5$ V feedback pulses requires a modification of this equation to:

$$V_{i+1} = V_i e^{-\frac{t_1}{\tau_{\mp 1V}}} e^{-\frac{t_1}{\tau_{\pm 0.5V}}} \pm c \qquad (10)$$

where $c = e^{-\frac{t_1}{\tau_{\mp 1V}}} e^{-\frac{t_1}{\tau_{\pm 0.5V}}} - 1.5 e^{-\frac{t_1}{\tau_{\pm 0.5V}}} + 0.5$ V, and $\tau_{\mp 1V}$ and $\tau_{\pm 0.5V}$ are the characteristic times during the $\mp 1$ V and $\pm 0.5$ V pulses, respectively.

In addition to this, the weight updates described before assume illumination of the synapse for the entire duration of the feedback pulse, or no illumination at all, shown by Figure S12a. By contrast, the (anti-)STDP updates depend on the time difference between the feedback spike and the illumination of the synapse ($\Delta t$), and allow varying degrees of overlap with the feedback spike, illustrated by Figure S12b. Figure 4 in the main text shows that the sign and magnitude of the weight changes depend on this delay time. Modifications to the previously derived equations to calculate weight updates as a function of $\Delta t$ are described below.

First, equation 8 is redefined to correct for any differences compared to the simple feedback pulse case. For the STDP measurement in Figure 4a in the main text, $W_{i+1}(V_{i+1}) = 17$ for $\Delta t = 22.5$ ms. This time delay corresponds to full overlap of the light pulse with the $-1$ V pulse, followed by a $+0.5$ V pulse in dark, as in the right-most condition in Figure S12b. The corresponding $V_{i+1}$ can therefore be calculated from equation 10, by setting $\tau_{-1V} = \tau_{light}$ and $\tau_{+0.5V} = \tau_{dark}$, yielding $V_{i+1} = -0.57$ V. Substitution of $W_{i+1}(V_{i+1})$ and the obtained value for $V_{i+1}$ into equation 7 gives:

$$W_i(V_i) = 1 - \frac{V_i}{0.57\text{ V}} \times 16 \tag{11}$$

We note that equation 8 gives a slightly different weight of $W_{i+1}(V_{i+1} = -0.57\text{ V}) = 21.5$. A possible explanation could be an imperfect overlap of the light pulse with the $-1$ V pulse, causing some overlap with the following $+0.5$ V pulse as well. Substitution of equation 10 into equation 11 to express the weight changes as a first-order linear recurrence relation gives:

$$W_{i+1} = e^{-\frac{t_1}{\tau_{\mp 1V}}} e^{-\frac{t_1}{\tau_{\pm 0.5V}}} W_i + d \tag{12}$$

where $d = 1 - e^{-\frac{t_1}{\tau_{\mp 1V}}} e^{-\frac{t_1}{\tau_{\pm 0.5V}}} \mp \frac{c}{0.57} \times 16$, and $c$ the term of equation 10.

In the derivation of equation 11, the $-1$ V pulse fully overlapped with the light pulse and the $+0.5$ V pulse was fully in dark, so $\tau_{-1V} = \tau_{light}$ and $\tau_{+0.5V} = \tau_{dark}$. However, other values of $\Delta t$ would give partial overlap with the voltage pulses. Figure 4 shows that this results in smaller modulation of the synaptic weight, which can be explained by a value for $\tau$ between $\tau_{light}$ and $\tau_{dark}$. Hence, instead of $\tau \in \{\tau_{light}, \tau_{dark}\}$, $\tau$ is a continuous function of delay $\Delta t$ for the (anti-)STDP updates. An expression for $\tau(\Delta t)$ can be found by setting equation 12 equal to the fitting equation of Figure 4:

$$e^{-\frac{t_1}{\tau_{\mp 1V}}} e^{-\frac{t_1}{\tau_{\pm 0.5V}}} W_1 + d = Ae^{\pm \Delta t/\tau_{fit}} \tag{13}$$

where $A$ and $\tau_{fit}$ are obtained from the empirical fits in Figure 4 and $W_1 = 1.0$ per the experimental conditions. To obtain $\tau(\Delta t)$, it is assumed that either the $\mp 1$ V or the $\pm 0.5$ V pulse is fully in dark, so $\tau_{\mp 1V} = \tau_{dark}$ or $\tau_{\pm 0.5V} = \tau_{dark}$. Based on this assumption, (partial) overlap of the light pulse with the $-1$ V part of the STDP pulse gives:

$$\tau_{-1V} = -\frac{t_1}{\ln\left[1.5 - \frac{0.5}{e^{-\frac{t_1}{\tau_{dark}}}} + \frac{0.57}{16e^{-\frac{t_1}{\tau_{dark}}}}\left(1 - 34.3e^{-\frac{\Delta t}{29.4\text{ ms}}}\right)\right]} \tag{14}$$

While overlap of the light pulse with the +0.5 V part of the STDP pulse gives:

$$\tau_{+0.5V} = -\frac{t_1}{\ln\left[\frac{1}{e^{-\frac{t_1}{\tau_{dark}}} - 1.5}\left(\frac{0.57}{16}\left(36.3e^{\frac{\Delta t}{11.3\text{ ms}}} + 1\right) - 0.5\right)\right]} \tag{15}$$

Similarly, for the anti-STDP pulses, ignoring the offset in the fitting equation, overlap of the light pulse with the +1 V pulse gives:

$$\tau_{+1V} = -\frac{t_1}{\ln\left[1.5 - \frac{0.5}{e^{-\frac{t_1}{\tau_{dark}}}} + \frac{0.57}{16e^{-\frac{t_1}{\tau_{dark}}}}\left(1 + 40.9e^{-\frac{\Delta t}{24.7\text{ ms}}}\right)\right]} \tag{16}$$

and overlap with the $-0.5$ V pulse yields:

$$\tau_{-0.5V} = -\frac{t_1}{\ln\left[\frac{1}{e^{-\frac{t_1}{\tau_{dark}}} - 1.5}\left(\frac{0.57}{16}\left(30.8e^{\frac{\Delta t}{17.5\text{ ms}}} - 1\right) - 0.5\right)\right]} \tag{17}$$

The equations derived above were validated by comparing weights obtained from simulations to the experimentally obtained weight changes in Figure 4. Figure S13a and b show a perfect match. Figure S13c shows a simple simulation of STDP weight updates of a single synapse. A random STDP voltage profile is applied to the synapse, which updates its weight depending on the overlap with the light pulse. Similar to the simulation in Figure S11c, this simple simulation shows the expected logarithmic weight increases, and exponential decays with each update. However, the STDP pulses allow both positive and negative weights for the same feedback pulse voltage profile depending on the overlap with the light pulse.

When the (anti-)STDP pulses are used to update the synapse array, $\Delta t$ is determined by searching $X(t)$ for a light pulse from $t = t_{spike} - 100$ ms to $t = t_{spike} + 100$ ms. If multiple light pulses are found, $\Delta t$ is determined from the light pulse that is closest in time to the feedback spike. If no light pulse is found, the synaptic weight does not change, in accordance with the experimental results in Figure 4. Currently, the timestep in the simulation and the light and voltage pulse durations are all 45 ms, per the experimental conditions in the main text. Consequently, the light pulses overlap fully with either the $-1$ V or the $+0.5$ V part of the STDP pulse, or not at all. This is a limitation of our current simulations that stems from the minimum step size of 45 ms in the measurements. Nevertheless, the (anti-)STDP updates can be extended easily to also allow the partial overlaps shown in Figure S12b. Smaller step sizes could be implemented with the currently derived equations, extrapolating our experimental results. In future work, the measurements could be repeated on a setup that allows smaller timesteps to validate that the equations and assumptions hold for shorter timescales as well. Equations 14 - 17 could then be updated accordingly if necessary.

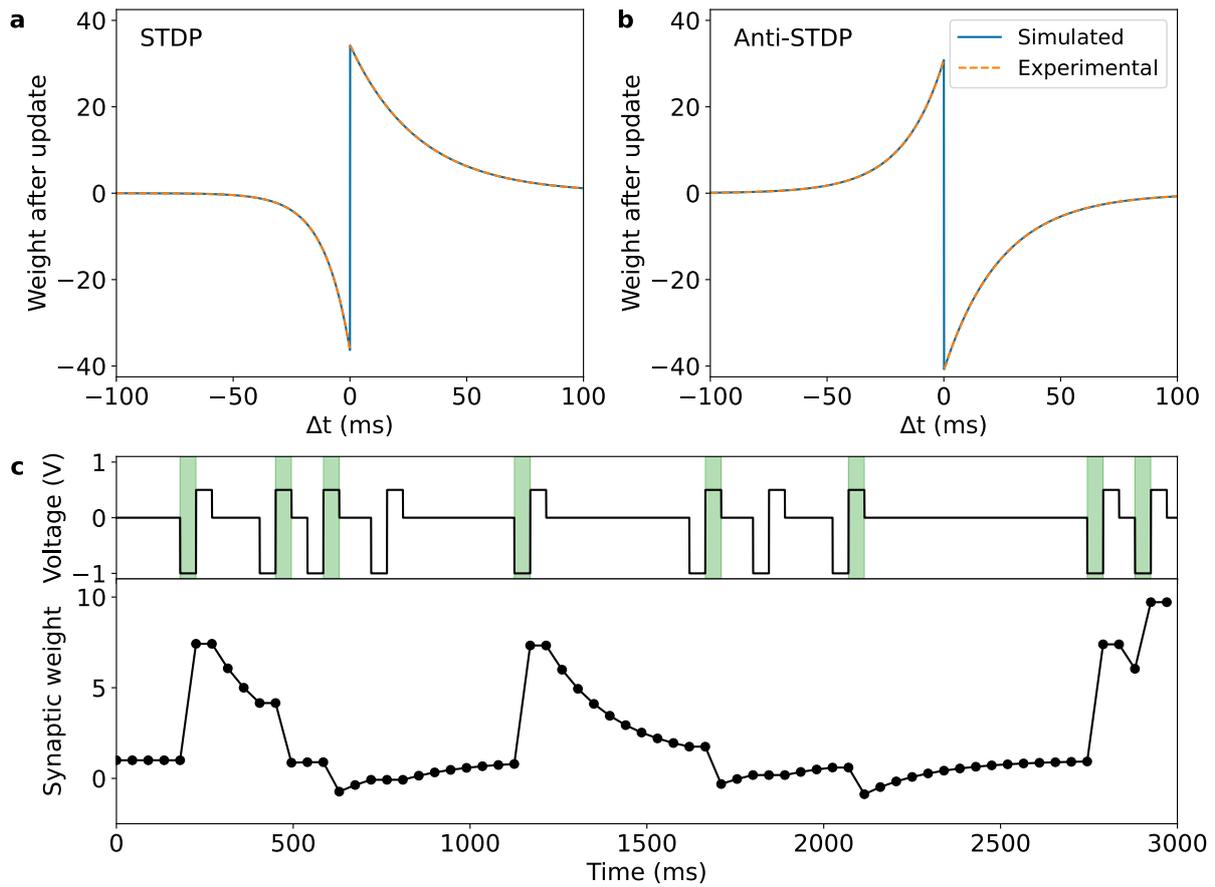

*Figure S13. STDP weight updates simulated for a single synapse to validate the simulation results. **(a)** Comparison of simulated Δt dependent weight updates for STDP with the experimental fit in Figure 4a. **(b)** Comparison of the simulated Δt dependent weight updates for anti-STDP with the experimental fit in Figure 4b. Simulated and experimentally obtained weight updates match perfectly for both update rules. **(c)** Simulation of STDP synaptic weight updates based on a random STDP voltage and light profile. The random STDP voltage profile is shown in the top panel. Green shaded regions indicate simultaneous illumination of the device. The synaptic weight in the bottom panel increases or decreases depending on what part of the voltage profile overlaps with the light pulse. Every marker represents the synaptic weight at a 45 ms timestep.*

# Supplementary Note 2. Attention-based learning with simple -1 V pulses

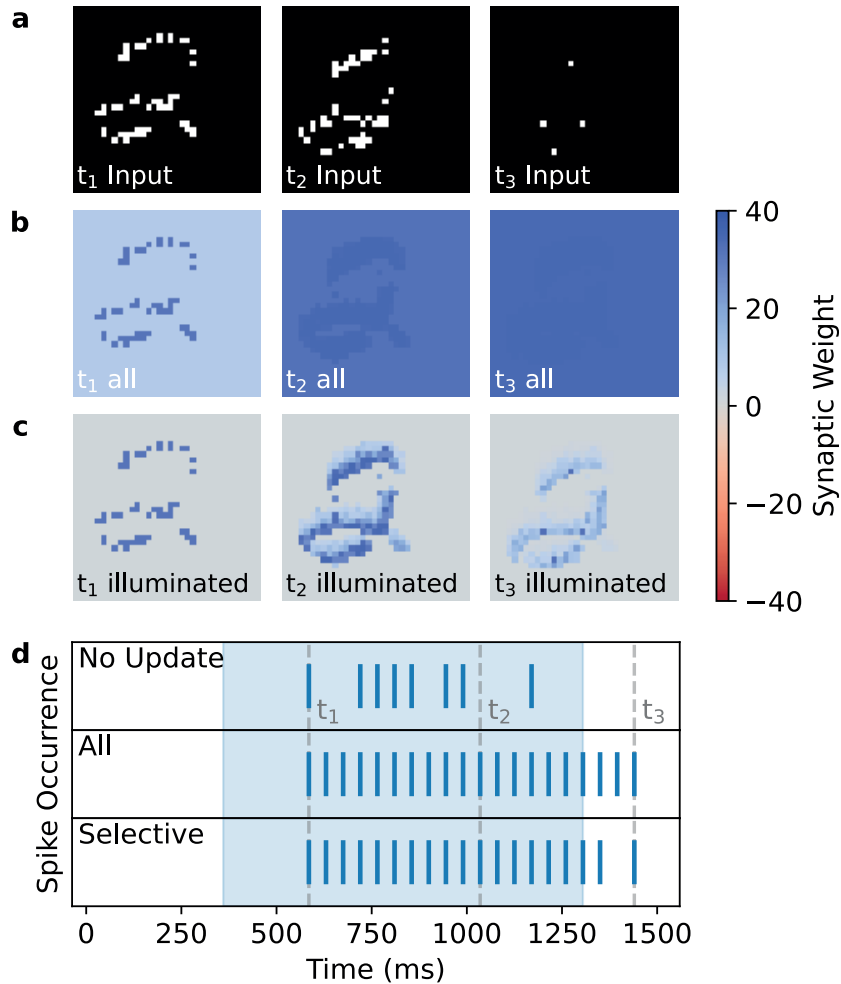

*Figure S14. The attention mechanism implemented with -1 V pulses. **(a)** The same input frames of the N-MNIST sample number 2 as in Figure 5 in the main text. The frames are taken at the same simulation times $t_1$ (585 ms), $t_2$ (1035 ms), and $t_3$ (1440 ms). **(b)** The synaptic weights of an optoelectronic synapse array to which -1 V pulses are applied to all synapses after each neuron spike. The three panels show the weights of the array after being presented the frames in (a). The synaptic weights of illuminated synapses are increased to a larger degree, but cumulative spiking causes all synapses to have the same weights after $t_3$. **(c)** The synaptic weights of an optoelectronic synapse array to which -1 V pulses are applied selectively only to illuminated synapses after each neuron spike. Larger synaptic weights are found only in regions that were illuminated with the number 2. **(d)** Event-plot of the neuron spikes over time. Higher spiking frequencies are found for the arrays implementing the weight changes in **(b)** ("All") and **(c)** ("Selective") compared to an array that does not implement any weight changes ("No Update"). The blue shaded region indicates the times during which the number 2 is visible in the N-MNIST frames, i.e. when the neuron should output spikes.*

Figure S14c and the corresponding Supplementary Movie demonstrate how a more top-down attention mechanism can be implemented by applying -1 V pulses only to illuminated synapses after each neuron spike. Positive weights are obtained only in the regions illuminated with the number 2, similar to the STDP-learning array in Figure 5b. However, applying the -1 V pulses to all pixels, as with the STDP-learning algorithm, causes all synaptic weights to converge to the same high value over time and does not lead to attention, as demonstrated by Figure S14b. These results illustrate that this top-down approach requires a more elaborate feedback mechanism compared to the (anti-

)STDP updates in Figure 5 in the main text. This makes the top-down approach more difficult to scale. Nevertheless, this approach could be worthwhile for more complex inputs where the (anti-)STDP updates cannot sufficiently separate features of interest.